\documentclass[lettersize,journal]{IEEEtran}
\usepackage{amsmath,amsfonts}
\usepackage{array}
\usepackage[caption=false,font=scriptsize,labelfont=sf,textfont=sf]{subfig}
\usepackage{hyperref}
\usepackage{textcomp}
\usepackage{stfloats}
\usepackage{url}
\usepackage{verbatim}
\usepackage{graphicx}
\usepackage{cite}
\usepackage[numbers]{natbib}
\usepackage{graphbox}
\usepackage{enumitem}
\usepackage{booktabs}
\usepackage{multirow}
\usepackage{makecell}
\usepackage{amssymb}
\usepackage[table,xcdraw,dvipsnames]{xcolor}
\usepackage{algorithm}
\usepackage{algpseudocode}
\usepackage{amsmath}
\usepackage{color}
\usepackage[normalem]{ulem}
\definecolor{burntorange}{HTML}{CC5500}

\begin{document}

\title{Transmission Interface Power Flow Adjustment: \\A Deep Reinforcement Learning Approach based on Multi-task Attribution Map}

\author{Shunyu~Liu,
        Wei~Luo,
        Yanzhen~Zhou,
        Kaixuan~Chen,
        Quan~Zhang,
        Huating~Xu,
        Qinglai~Guo,
        Mingli~Song
        % <-this % stops a space
        \thanks{This article has been accepted for publication by IEEE Transactions on Power Systems. The published version is available at \url{https://doi.org/10.1109/TPWRS.2023.3298007}.
        This work was supported in part by the National Key R\&D Program of China under Grant 2018AAA0101503, and in part by the Science and Technology Project of SGCC (State Grid Corporation of China): Fundamental Theory of Human-in-the-Loop Hybrid-Augmented Intelligence for Power Grid Dispatch and Control. 
        \textit{(Shunyu~Liu and Wei~Luo contributed equally to this work.)} \textit{(Corresponding author: Mingli~Song.)}
        }
        \thanks{Shunyu Liu, Wei Luo, Kaixuan Chen, and Mingli Song are with the College of Computer Science and Technology, Zhejiang University, Hangzhou 310027, China (e-mail: liushunyu@zju.edu.cn; \text{davidluo}@zju.edu.cn; chenkx@zju.edu.cn; brooksong@zju.edu.cn).}
        \thanks{Yanzhen Zhou and Qinglai Guo are with the Department of Electrical Engineering, Tsinghua University, Beijing 100084, China (e-mail: \text{zhouyzh}@mail.tsinghua.edu.cn; guoqinglai@tsinghua.edu.cn).}
        \thanks{Quan Zhang and Huating Xu are with the College of Electrical Engineering, Zhejiang University, Hangzhou 310027, China (e-mail: \text{quanzzhang}@zju.edu.cn; xu\_huating@zju.edu.cn).}
        \thanks{Digital Object Identifier 10.1109/TPWRS.2023.3298007}
}

\markboth{IEEE TRANSACTIONS ON POWER SYSTEMS}%
{Liu \MakeLowercase{\textit{et al.}}: Transmission Interface Power Flow Adjustment: A Deep Reinforcement Learning Approach based on Multi-task Attribution Map}

\IEEEpubid{\begin{minipage}{\textwidth}\ \\[30pt] \centering 0885-8950~\copyright~2023 IEEE. Personal use is permitted, but republication/redistribution requires IEEE permission.\\See https://www.ieee.org/publications/rights/index.html for more information.\end{minipage}}


\maketitle

\begin{abstract} 
  Transmission interface power flow adjustment is a critical measure to ensure the security and economy operation of power systems. However, conventional model-based adjustment schemes are limited by the increasing variations and uncertainties occur in power systems, where the adjustment problems of different transmission interfaces are often treated as several independent tasks, ignoring their coupling relationship and even leading to conflict decisions. In this paper, we introduce a novel data-driven deep reinforcement learning~(DRL) approach, to handle multiple power flow adjustment tasks jointly instead of learning each task from scratch. At the heart of the proposed method is a multi-task attribution map~(MAM), which enables the DRL agent to explicitly attribute each transmission interface task to different power system nodes with task-adaptive attention weights. Based on this MAM, the agent can further provide effective strategies to solve the multi-task adjustment problem with a near-optimal operation cost. Simulation results on the IEEE 118-bus system, a realistic 300-bus system in China, and a very large European system with 9241 buses demonstrate that the proposed method significantly improves the performance compared with several baseline methods, and exhibits high interpretability with the learnable MAM.
\end{abstract}


\section{Introduction}
\IEEEPARstart{P}{ower} systems are complex nonlinear physical systems with high uncertainty~\cite{hatziargyriou2020definition}. With the rapid expansion of the scale of power systems and the increasing imbalance of power demand and generation, the problems of power systems, such as security and economy, become much more challenging~\cite{chen2022reinforcement}. To monitor the operation state of power systems with massive variables, operators tend to take the transmission interface into consideration instead of a single element. A specific transmission interface is composed of a set of transmission lines with the same direction of active power flow and close electrical distance~\cite{zhang2006study,kaymaz2007transmission}. Operators can analyze and control the operation state of the power system by monitoring the power flow of different transmission interfaces. The total transfer capability of critical transmission interfaces is widely used in practice to provide power system security margins~\cite{min2006total,sun2015automatic,liu2015iterative}. Once the power flow of a critical transmission interface is overloaded, it inevitably poses a great threat to the power system and even leads to the emergence of cascading blackouts. In this way, the power flow through a transmission interface is typically regulated at a pre-scheduled range, thereby ensuring the stability and reliability of power system operation~\cite{lin2019tie,mudaliyar2020distributed}. 

Transmission interface power flow adjustment serves as an important defensive means to satisfy this pre-scheduled security constraint. To realize power flow adjustment for different transmission interfaces, a direct approach is generation dispatch. Conventional dispatch methods are highly dependent on the system model, which can be mainly categorized into two classes: optimal power flow-based methods~\cite{capitanescu2011state} and sensitivity analysis-based ones~\cite{nguyen2003dynamic,hiskens2006sensitivity,dutta2008optimal}. Optimal power flow-based methods mainly rely on numerical optimization that transforms the dispatch problem into a constrained programming problem, while sensitivity analysis-based methods iteratively calculate the sensitivity indexes to determine the candidate generators. The conventional methods are based on mathematical models that represent the power system, where the models are subject to certain assumptions and simplification. As a result, these methods may not be able to capture all the complex dynamics that occur in a real power system. Moreover, this model-based mechanism greatly suffers from a heavy computational burden with the growing scale of the power system~\cite{molzahn2017survey}. The power flow constraints of different transmission interfaces are closely intertwined and even conflict with each other in extreme scenarios, which significantly limits the feasible solution space. Thus, due to the combinatorial explosion and complicated constraints in the nonlinear nonconvex searching space, the conventional methods may easily fail to find an optimal solution and are no longer acceptable.

\IEEEpubidadjcol

To alleviate these issue, data-driven framework based on deep reinforcement learning~(DRL) has been proposed as a competent alternative~\cite{zhang2019deep}. As a solution to control problems, model-free DRL exploits the large capacity of deep neural networks to extract effective features from input states and deliver response actions in an end-to-end fashion~\cite{books:books/lib/SuttonB98}. Recently, many studies have shown the remarkable potential of this learning paradigm in many power system control problems, including voltage control~\cite{DuanJiajun,yang2019two,cao2021deep}, economic dispatch~\cite{liu2018distributed,dai2019distributed,li2021virtual} and emergency control~\cite{huang2019adaptive,chen2020model}. DRL can directly learn from high-dimension power system data without being constrained by a fixed model, providing a more robust and adaptive control strategy under various scenarios. Moreover, DRL can realize efficient inference based on neural networks, making it more suitable for real-world applications with high demands for rapid response~\cite{cao2020reinforcement}. For transmission interface power flow adjustment problem, reference~\cite{gao2021hybrid} proposes to use Proximal Policy Optimization~(PPO) to train the DRL policy. To avoid the conflict problem of training scenarios with distinct patterns, this method first clusters the training scenarios and then employs multiple DRL agents for each cluster. Despite the promising results achieved, existing works are still limited by individually training policies for each specific transmission interface task, which requires a large amount of interaction data and ignores the coupling relationship across various tasks.

It is inevitable to monitor multiple transmission interfaces simultaneously, while each adjustment task has its own objective function. Therefore, directly training the policy network for multiple tasks in a simple parameter-sharing manner is not feasible. Despite the existence of commonalities, the differences between tasks may require network parameters to be updated in opposite directions, resulting in the optimization issue of gradient conflict~\cite{yu2020gradient}. To remedy the conflict issue and simplify the exploration space, disentangling the relationship between multiple tasks is especially critical. On the other hand, different adjustment tasks under the same power system topology share the same operation state space and dispatch action space. Thus, it is beneficial to learn multiple tasks jointly instead of learning them from scratch separately, with the aim of leveraging the shareable representation ability and decision-making pattern. A single policy network trained on multiple tasks jointly would be able to generalize its knowledge, which further leads to improved efficiency~and~performance.

In this paper, we introduce a novel multi-task reinforcement learning method, to learn multiple power flow adjustment tasks jointly. Unlike traditional methods that merely share network parameters across tasks in an intertwined manner, the proposed method performs a selective feature exaction for each transmission interface task in an attribution manner. Specifically, we design a multi-task attribution map~(MAM) to explicitly distinguish the impact of different power system nodes on each transmission interface task. To construct the MAM, we first learn the task representations based on the common task encoder to capture the relationship among tasks. Meanwhile, the graph convolution network is used to learn the node features of the power system operation state. Then each task representation serves as a query value to calculate the node-level attention weights, resulting in a learnable MAM. The MAM enables the model to mitigate the gradient conflict issue. It selectively integrates the underlying node features, achieving a compact state representation. Finally, we further use Dueling Deep Q Network to derive the final reinforcement learning policy, realizing effective power flow adjustment. 

Our main contributions are summarized as follows:
\begin{itemize}
  \item This work is therefore the first dedicated attempt towards learning multiple transmission interface power flow adjustment tasks jointly, a highly practical problem yet largely overlooked by existing literature in the field of the power system.
  \item We design a novel deep reinforcement learning~(DRL) method based on multi-task attribution map~(MAM) to handle multiple adjustment tasks jointly, where MAM enables the DRL agent to selectively integrate the node features into a compact task-adaptive representation for the final adjustment policy.
  \item Simulations are conducted on the IEEE 118-bus system, a realistic 300-bus system in China, and a very large European 9241-bus system, demonstrating that the proposed method brings remarkable improvements to the existing methods. Moreover, we verify the interpretability of the learnable MAM in different operation scenarios.
\end{itemize}

\section{Problem Formulation\label{sec:problem}}

For the transmission interface power flow adjustment problem, the adjustment objective is to bring the power flow of the transmission interface to a pre-scheduled range by generation dispatch. Specifically, we assume that there are $N_{\Phi}$ transmission interfaces $\{\Phi_1,\Phi_2,\cdots,\Phi_{N_{\Phi}}\}$ in a given power system. Each transmission interface $\Phi$ is a set of several transmission lines with the same direction of active power flow and close electrical distance. The power flow of the transmission interface $\Phi$ is defined as $P_{\Phi} = \sum_{\ell\in\Phi}{P_\ell}$, where $P_\ell$ is the active power of the transmission line $\ell$. Generation dispatch is utilized to adjust the power flow of the transmission interface. The power flow of the power system after generation dispatch is solved by the equations:
\begin{align}
  &\begin{aligned}
  P^{G}_{i}-P^{L}_{i}=|V_{i}| \sum_{j=1}^{N_{B}} |V_{j}|\left(G_{i j} \cos\omega_{ij} + B_{i j} \sin\omega_{ij}\right),\end{aligned} \\
  &\begin{aligned}Q^{G}_{i}-Q^{L}_{i}=|V_{i}| \sum_{j=1}^{N_{B}} |V_{j}|\left(G_{i j} \sin \omega_{ij}-B_{i j} \cos \omega_{ij}\right),\end{aligned}
\end{align}
where $P^{L}_{i}$ and $Q^{L}_{i}$ are the active and reactive power consumption of the load at the bus $i$ respectively, $P^{G}_{i}$ and $Q^{G}_{i}$ are the active and reactive power production of the generator at the bus $i$ respectively, $V_i$ is the voltage magnitude of bus $i$, $\omega_{ij}=\omega_{i}-\omega_{j}$  is the phase difference between bus $i$ and $j$, $G_{ij}$ and $B_{ij}$ are the conductance and susceptance matrix between bus $i$ and $j$, $N_{B}$ is the total number of buses, $N_{L}$ is the number of loads and $N_{G}$ is the number of generators. To satisfy safety margin constraints under N-1 contingency conditions, the pre-scheduled range of each transmission interface $\Phi$ is given as $[\sigma_\Phi^{-}, \sigma_\Phi^+]$. $\sigma_\Phi^{-}$ and $\sigma_\Phi^+$ represent the upper and lower limit of the transmission interface total transfer capability~\cite{min2006total,sun2015automatic,liu2015iterative}, respectively. We need to adjust the power flow of the transmission interface to the pre-scheduled range via generation dispatch to ensure the safe operation of the entire power system.

\subsection{Task Setting} 
In this paper, we focus on training \emph{one} multi-task policy to learn \emph{multiple} tasks jointly instead of training different single-task policies for each task separately. Moreover, we consider two types of tasks in the transmission interface power flow adjustment problem: 
\begin{itemize}[leftmargin=*]
  \item \textbf{Single-Interface Task.} Each task $\mathcal{T}$ consists of the adjustment problem of \emph{one} transmission interface $\Phi$, \emph{e.g.} $\mathcal{T}(\Phi)$.
  \item \textbf{Multi-Interface Task.} Each task $\mathcal{T}$ consists of the adjustment problems of \emph{$M$} transmission interfaces, \emph{e.g.} $\mathcal{T}(\{\Phi_m\}_{m=1}^{M})$, where $M$ satisfies $ 1 < M < N_{\Phi}$.
\end{itemize}
In our simulation, each test scenario has only one specified task. In a given scenario with the single-interface task, the multi-task policy only needs to adjust the power flow for one transmission interface. Moreover, the same multi-task policy can also deal with other scenarios with different single-interface tasks. In contrast, for a scenario with the multi-interface task, the multi-task policy need to adjust the power flow of multiple transmission interfaces to their pre-scheduled ranges at the same time.

\subsection{Markov Decision Process} 

The transmission interface power flow adjustment task $\mathcal{T}$ can be formalized as a Markov decision process (MDP). An MDP is represented by a tuple $\langle \mathcal{S,A},\mathcal{P},\mathcal{R}, \gamma\rangle $. $\mathcal{S}$ is the set of continuous states, and $\mathcal{A}$ is the set of discrete actions. $\mathcal{P}$ is the state transition function, $\mathcal{R}$ is the reward function, and $\gamma \in [0,1]$ is the discount rate. The multi-step element in the power flow adjustment problem is the dispatch action sequence, where the maximum sequence length is set as 50. At each control step $t$, the agent with a given state $s \in \mathcal{S}$ can select the action $a \in \mathcal{A}$ through its policy $\pi(a|s)$. Then the agent will receive a reward $r$ and the next state of the environment $s'$ according to the reward function $\mathcal{R}(s,a)$ and the transition function $\mathcal{P}(s' | s, a)$. For an agent, its goal is to find the optimal policy $\pi_{\theta}$ parameterized by $\theta$ to maximize the discounted return:
\begin{align}
J^{\pi}(\theta) = \mathbb{E}_{s\sim \rho(\cdot), \pi}\left[\sum_{t=0}^{\infty}{\gamma^{t} \mathcal{R}(s_{t},a_{t})} \mid s_0 = s \right], 
\label{eq:single}
\end{align}
where $\rho(\cdot)$ is the initial state distribution and $a_t \sim \pi_{\theta}(a_t|s_t)$. Moreover, the goal of multi-task reinforcement learning is to learn a compact policy $\pi$ for various tasks drawn from a distribution of tasks $p(\mathcal{T})$. We define a task-conditioned policy $\pi(a | s,z)$, where $z$ denotes a task representation for each task $\mathcal{T}$. The task representation $z$ can be provided as a one-hot vector or in any other form. Each task $\mathcal{T}$ is a standard MDP. Considering the average expected return across all tasks sampled from $p(\mathcal{T})$, the multi-task policy $\pi$ parameterized by $\theta$ can be optimized according to maximize the objective:
\begin{align}
  J^{\pi}(\theta) = \mathbb{E}_{\mathcal{T}\sim p(\mathcal{T})}\left[ J^{\pi}_\mathcal{T}(\theta) \right],
\end{align}
where $J^{\pi}_\mathcal{T}(\theta)$ is obtained directly from Eq.~\ref{eq:single} with task $\mathcal{T}$. The detailed MDP formulation of the power system is defined as:

\textbf{State.} 
We define the power system operation state as a graph $s = (A,F)$, where $A$ is the adjacency matrix and $F$ is the node feature matrix. The graph-based state is modeled from the power system by considering the bus nodes as vertices and transmission lines as edges. For each bus node $B$, its feature vector can be described as $F_B = [P_{B}, Q_{B}, V_{B}, \omega_B]^T$, where $P_{B}$ and $Q_{B}$ are the active and reactive power of the bus node $B$ respectively, $V_{B}$ and $\omega_B$ is the voltage magnitude and phase of the bus node $B$ respectively. We use power flow calculation to get the next state based on the current power system operation state with the current dispatch action.

\textbf{Action.} 
The action adopted in our setting is generation dispatch, which enables the agent to change the active power of the controllable generators. The action set is defined as $\mathcal{A} = \mathcal{G} \times \{c_G^-, c_G^+\}$, where $\mathcal{G}$ is the set of the controllable generators. $\sigma_G^-$ and $\sigma_G^+$ represent the increasing and decreasing rates of the active power, respectively. We set $\sigma_G^-$ as $90\%$ and $ \sigma_G^+$ as $110\%$ based on limits of the ramp rates. At each discrete time step, the agent can reduce the active power of a selected generator to 90\%-level or increase to 110\%-level. If the current generation is already at the upper or lower limit of the generator capacity, the corresponding invalid dispatch action will be masked. It is worth mentioning that slack generators are not included in controllable generators.

\textbf{Reward.} Considering the both power flow adjustment and economic dispatch cost, the total reward function is defined as
\begin{align}
  \mathcal{R}_{tot}\left(s,a\right) = \mathcal{R}_{pf}\left(s,a\right) + \mathcal{R}_{ed}\left(s,a\right). 
\end{align}
To satisfy the limits of transmission interface power flow, the reward function of the single-interface task with the transmission interface $\Phi$ is defined as 
\begin{align}
  \mathcal{R}_{pf}\left(s,a;\Phi\right) = -\left| P_{\Phi} - \frac{\sigma_{\Phi}^- + \sigma_{\Phi}^+}{2}   \right|. 
\end{align}
For the multi-interface task with $\{\Phi_m\}_{m=1}^M$, we focus on the worst case and define the reward function as 
\begin{align}
  \mathcal{R}_{pf}\left(s,a;\{\Phi_i\}_{m=1}^M\right) = \min_{m \in \left\{1,2,\cdots,M\right\} } \left\{ \mathcal{R}_{pf}\left(s,a;\Phi_m\right) \right\}.
\end{align}
The most commonly used reward assumption of the economic dispatch is quadratic:
\begin{align}
  \mathcal{R}_{ed}\left(s,a\right) = -\sum_{i=1}^{N_G} \left(\alpha_i (P_i^G)^2 + \beta_i P_i^G + \lambda_i \right),
\end{align}
where $\alpha_i$, $\beta_i$, and $\lambda_i$ are the generation cost coefficients of the generator $i$. $P^G_i$ is the active power production of the generator $i$.
Moreover, the reward is directly set to $\mathcal{R}_{tot}\left(s,a\right)=-100$ to penalize the divergence of power flow and the overload of slack generators, and $\mathcal{R}_{tot}\left(s,a\right)=100$ to incentive the success of adjustment where the power flow of all adjusted transmission interfaces is within the pre-sheduled range.

\begin{figure*}[!t]
  \centering
  \includegraphics[scale=1.0]{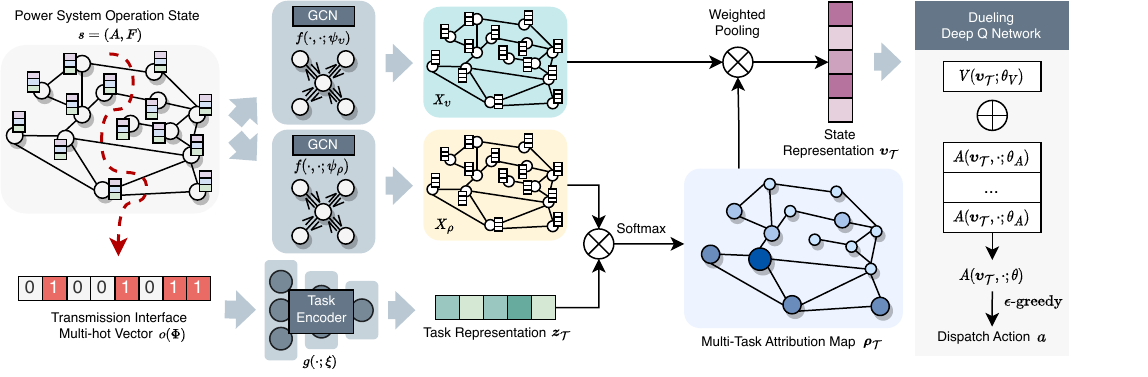}
  \caption{Illustration of the proposed method.}
  \label{fig:framework}
\end{figure*}

\section{Methodology\label{sec:method}}

In what follows, we detail the proposed method for multi-task transmission interface power flow adjustment. As shown in Figure~\ref{fig:framework}, we first use the graph convolution network to learn the node features of power system operation state. Meanwhile, the task representations are obtained based on the common task encoder to capture the relationship among tasks. Then each task representation acts as a query value to calculate the node-level attention weights based on the node features, resulting in the task-adaptive attribution map. The attribution map enables the model to selectively integrates the underlying node features, achieving a compact state representation. Finally, we further use Dueling Deep Q Network to derive the final reinforcement learning policy. 

\subsection{Graph Convolution Network}

A power system operation state $s \in \mathcal{S}$ is represented as $(A, F)$, where $A \in \{0,1\}^{N \times N}$ is the adjacency matrix with $N$ nodes and $A_{ij} = 1$ indicates that there is a transmission line between node $i$ and $j$, and $F \in \mathbb{R}^{N \times d_{s}}$ is the node feature matrix assuming each node has $d_{s}$ features. To further extract information from graph-based data, our framework uses Graph Convolution Network~(GCN)~\cite{GCN} to obtain node embeddings layer by layer. GCN enables the nodes to exchange information with their neighbors in each convolutional layer, followed by applying learnable filters and some non-linear transformation. As depicted in the left column of Figure~\ref{fig:framework}, the framework first uses GCN $f(\,\cdot\,;{\psi}):\mathbb{R}^{d_s}\to\mathbb{R}^{d_x}$ parameterized by $\psi$ to extract the node-level embeddings with dimension $d_x$ following the general message passing mechanism as
\begin{align}
  H^{(k)} = \operatorname{ReLU}\left(\hat{D}^{-\frac{1}{2}} \hat{A} \hat{D}^{-\frac{1}{2}} H^{(k-1)} W_{\psi}^{(k-1)}\right),
  \label{eq:gcn}
\end{align}
where $\hat{A} = A + I$ is the adjacency matrix with added self-connections. $I$ is the identity matrix. $\hat{D} = \sum_{j}\hat{A}_{ij}$ is the degree matrix. $W_\psi^{(k)} \in \mathbb{R}^{d^{(k)} \times d^{(k+1)}}$ is a trainable weight matrix with parameters $\psi$. $\operatorname{ReLU}(\cdot) = \max (0,\cdot)$ denotes an activation function. $H^{(k)} \in \mathbb{R}^{N \times d^{(k)}}$ are the embedding matrix with dimension $d^{(k)}$ computed after $k$ steps of the graph convolution network. The input node embeddings $H^{(0)}$ is initialized using the node feature matrix $F$. After running $K$ iterations of Eq.~(\ref{eq:gcn}), the graph convolution network can generate the final node-level embedding matrix as
$X = H^{(K)} \in \mathbb{R}^{N \times d_{x}}$.
Here we adopt two GCN branches to obtain two node-level embedding matrices as shown in Figure~\ref{fig:framework}:  
\begin{align}
X_\rho    = f(A, F;{\psi_\rho   }) \in \mathbb{R}^{N\times d_x},
\end{align}
\begin{align}
X_\upsilon   = f(A, F;{\psi_\upsilon }) \in \mathbb{R}^{N\times d_x},
\end{align}
where $X_\rho$ is used for generating the multi-task attribution map. $X_\upsilon$ is weighted pooled by this attribution map to give a compact state representation.

\subsection{Multi-Task Attribution Map}

Our main idea in achieving knowledge transfer among multiple tasks is to learn a multi-task attribution map to realize task-adaptive node attribution. To generate the multi-task attribution map, we first need to capture and exploit both the common structure and the unique characteristics of tasks by learning task representation. We associate each task $\mathcal{T}$ with a representation $\boldsymbol{z}_\mathcal{T} \in \mathbb{R}^{d_x}$ and expect it to reflect different properties of tasks. For modeling task similarity, all tasks share a common representation encoder, which takes as input the task vector and outputs the task representation. The task encoder is trained on all tasks. 

Concretely, the task encoder is implemented by a Multi-layer Perceptron~(MLP) network $g(\,\cdot\, ;\xi )$ parameterized by $\xi$. Considering the single-interface task $\mathcal{T}(\Phi)$, the task representation $z_\mathcal{T}$ is calculated by
\begin{align} 
  \boldsymbol{z}_\mathcal{T} = g(o(\Phi) ;\xi ) \in \mathbb{R}^{d_x},
\end{align}
where $o(\Phi) = \left[o(\Phi)_1, o(\Phi)_2,\cdots,o(\Phi)_{N_\ell}\right]^T \in \{0,1\}^{N_\ell}$ is the transmission interface multi-hot vector, indicating whether the transmission line $\ell$ belongs to the transmission interface $\Phi$. $N_\ell$ is the total number of transmission lines. $o(\Phi)_i = 1$ if $\ell_i \in \Phi$ and $o(\Phi)_i = 0$ if $\ell_i \notin \Phi$. Moreover, the task representation $z_\mathcal{T}$ for the multi-interface task $\mathcal{T}(\{\Phi_m\}_{m=1}^M)$ is calculated by
\begin{align}
  \boldsymbol{z}_\mathcal{T} = \frac{\sum_{m=1}^M g(o(\Phi_m) ;\xi )}{M} \in \mathbb{R}^{d_x}.
\end{align}
When there is no task encoder but only the original vector is used to represent the tasks, each task is usually orthogonal independent. The task encoder allows us to express the relationship between tasks in the representation space, which enables agents to use the similarity between tasks to generalize their knowledge.

Furthermore, to obtain the attribution map $\rho$, we use the task representation $\boldsymbol{z}_\mathcal{T}$ as a query to calculate the novel-level attention weight:
\begin{align}
  \boldsymbol{\rho }_\mathcal{T}   = \operatorname{softmax}( X_\rho  \boldsymbol{z}_\mathcal{T}) \in \mathbb{R}^{N}.
\end{align}
This attribution map $\rho$ is task-adaptive based on the task representation $\boldsymbol{z}_\mathcal{T}$. With the attribution map $\rho$, the agent can explicitly distinguish the impact of different power system nodes on each adjustment task. Then the compact state representation $\boldsymbol{\upsilon}_\mathcal{T}$ is obtained via weighted pooling operation as follows:
\begin{align}
  \boldsymbol{\upsilon}_\mathcal{T}  = X_\upsilon \boldsymbol{\rho }_\mathcal{T} \in \mathbb{R}^{d_x}.
\end{align}
The state representation $\boldsymbol{\upsilon}_\mathcal{T}$ selectively integrates the underlying node-level embeddings, which enables the agent to further derive a policy to handle the task $\mathcal{T}$.

\subsection{Dueling Deep Q Network}

We use a value-based method to calculate the final reinforcement learning policy. The value-based method tends to assess the quality of a policy $\pi$ by the action-value function $Q^{\pi}$ defined as:
\begin{align}
    Q^\pi(s,a)=\mathbb{E}_{\pi}\left[\sum_{t=0}^{\infty}{\gamma^t \mathcal{R}(s_{t},a_{t})} | s_0=s,a_0=a\right],
\end{align}
which denotes the expected discounted return after the agent executes an action $a$ at state $s$.
A policy $\pi^*$ is optimal if:
\begin{align}
Q^{\pi^*}(s,a) \geq Q^\pi(s,a), \forall{ \pi, s \in \mathcal{S}, a \in \mathcal{A}}.
\end{align}
There is always at least one policy that is better than or equal to all other policies~\cite{books:books/lib/SuttonB98}. All optimal policies share the same optimal action-value function defined as $Q^*$. It is easy to show that $Q^*$ satisfies the Bellman optimality equation:
\begin{align}
    Q^*(s,a) = \mathbb{E}_{s' \sim \mathcal{P}(\cdot |s,a)}\left[\mathcal{R}(s,a)+\gamma \max_{a'\in\mathcal{A}}Q^*(s',a')\right].
\end{align}

To estimate the optimal action-value function $Q^*$, Deep $Q$-Networks~(DQN)~\cite{DQN15} uses a neural network $Q(s,a; \theta)$ with parameters $\theta$ as an approximator. Here we take the state representation $\boldsymbol{\upsilon}_\mathcal{T}$ as input and adopt a dueling network architecture~\cite{DuelingDQN}, explicitly separates the estimation of state values and state-dependent action advantages. This corresponds to the following factorization of action values:
\begin{align}
  \begin{split}
  Q(\boldsymbol{\upsilon}_\mathcal{T}&,a; \theta) = \\& V(\boldsymbol{\upsilon}_\mathcal{T} ; \theta_V) + A(\boldsymbol{\upsilon}_\mathcal{T},a;\theta_A) - \frac{\sum_{a'\in\mathcal{A}}A(\boldsymbol{\upsilon}_\mathcal{T},a';\theta_A)}{|\mathcal{A}|},
\end{split}
\end{align}
where $V(\,\cdot\,; \theta_V):\mathbb{R}^{d_x}\to\mathbb{R}$ is the state value network and $A(\,\cdot\,;\theta_A):\mathbb{R}^{d_x}\to\mathbb{R}^{|\mathcal{A}|}$ is the action advantage network. For transmission interface power flow adjustment, most dispatch actions have little impact on the power system. The dueling architecture enable the agent to learn the individual value of the state itself, without having to learn the utility of each action on the state. This is especially useful in states where the actions have no effect on the environment~\cite{DuelingDQN}. Finally, the dispatch action can be obtained by a greedy policy $\pi_\mathcal{T}(s) = \arg\max_{a\in \mathcal{A}} Q(\boldsymbol{\upsilon}_\mathcal{T},a;\theta)$.

We optimize the networks by minimising the following temporal-difference~(TD) loss based on the Bellman equation:
\begin{align}
    \mathcal{L}(\psi,\xi ,\theta)=\mathbb{E}_{(\mathcal{T},s,a,r,s') \sim \mathcal{D}}\left[\left(y-Q(\boldsymbol{\upsilon}_\mathcal{T},a;\theta)\right)^2\right],
\end{align}
where $\mathcal{D}$ is the replay buffer of the transitions, 
$y = r+\gamma Q(\boldsymbol{\upsilon}_\mathcal{T}',\mathop{\arg\max}_{a'}Q(\boldsymbol{\upsilon}_\mathcal{T}',a';\theta);\theta^-)$
and $\theta^-$ represents the parameters of the target network~\cite{DoubleDQN}. An algorithmic description of the training procedure is given in Algorithm~\ref{alg:learn}.

\algdef{SE}[SUBALG]{Indent}{EndIndent}{}{\algorithmicend\ }%
\algtext*{Indent}
\algtext*{EndIndent}
\algnewcommand{\LineComment}[1]{\State \textcolor{gray}{\(\triangleright\) #1}}

\begin{algorithm}[!t]
	\caption{Learning Algorithm for MAM}
	\label{alg:learn}
  \begin{flushleft}
    \textbf{Initialize:} Two GCN Branches $f(\,\cdot\,;\psi_\rho)$ and $f(\,\cdot\,;\psi_\upsilon)$, Task Encoder $g(\,\cdot\,;\xi)$, DQN $Q(\,\cdot\,;\theta)$, Replay Buffer $\mathcal{D}$

    \textbf{Return:} Optimized Network Parameters $\{\psi, \xi, \theta\}$
  \end{flushleft}
	\begin{algorithmic}[1]
    \Repeat
      \State Sample a test scenario with the task from $p(\mathcal{T})$
      \State Obtain the initial state $s_0=(A,F_0)$
      \While{not terminal}
        \LineComment{Construct the multi-task attribution map}
        \State Calculate the node embedding matrices $X_\rho,X_\upsilon$
        \State Calculate the task representation $\boldsymbol{z}_\mathcal{T}$
        \State Construct the attribution map $\boldsymbol{\rho}_\mathcal{T}$
        \LineComment{Deliver the final policy}
        \State Calculate the state representation $\boldsymbol{\upsilon}_\mathcal{T}$
        \State Calculate the action values $Q(\boldsymbol{\upsilon}_\mathcal{T},\cdot;\theta)$
        \State Execute the action $a_t = \mathop{\arg\max}_a Q(\boldsymbol{\upsilon}_\mathcal{T},a;\theta)$
        \State Receive the reward $r_{t+1}$ and the next state $s_{t+1}$
        \State Store the sample to the buffer $\mathcal{D}$
        \LineComment{Train the networks}
        \State Sample the random minibatch from the buffer $\mathcal{D}$
        \State Compute the TD loss $\mathcal{L}(\psi,\xi ,\theta)$
        \State Perform gradient descent on the TD loss $\mathcal{L}(\psi,\xi ,\theta)$
      \EndWhile
    \Until converge
	\end{algorithmic}
\end{algorithm}

\section{Case Study\label{sec:case}}

To demonstrate the effectiveness of the proposed \emph{Multi-task Attribution Map}~(MAM) method, case studies are conducted based on two small-scale systems and one large-scale system using the PandaPower simulation~\cite{pandapower}. In this section, we first provide the details for the scenario generation process. Then the compared methods and parameter settings are introduced. The comparison results of different baselines are reported to evaluate the performance of MAM. Moreover, several visualization examples of MAM in different scenarios are given to study the internal mechanism of MAM.

\begin{figure}[!b]
  \centering
  \includegraphics[width=0.48\textwidth]{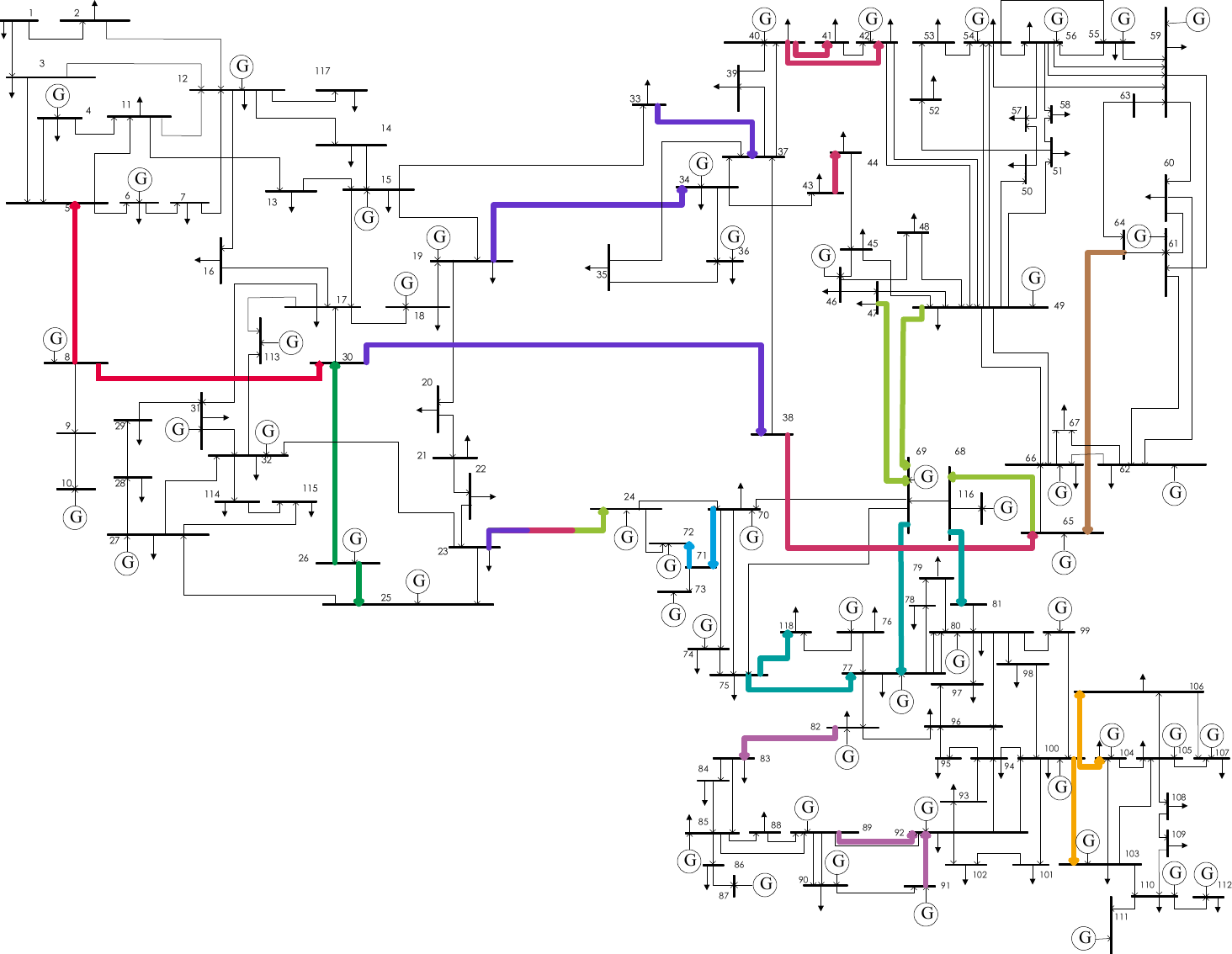}
  \caption{Illustration of the IEEE 118-bus system. Different transmission interfaces are represented by different colors. Please zoom for better view.}
  \label{fig:case118}
\end{figure}

\begin{figure}[!b]
  \centering
  \includegraphics[width=0.48\textwidth]{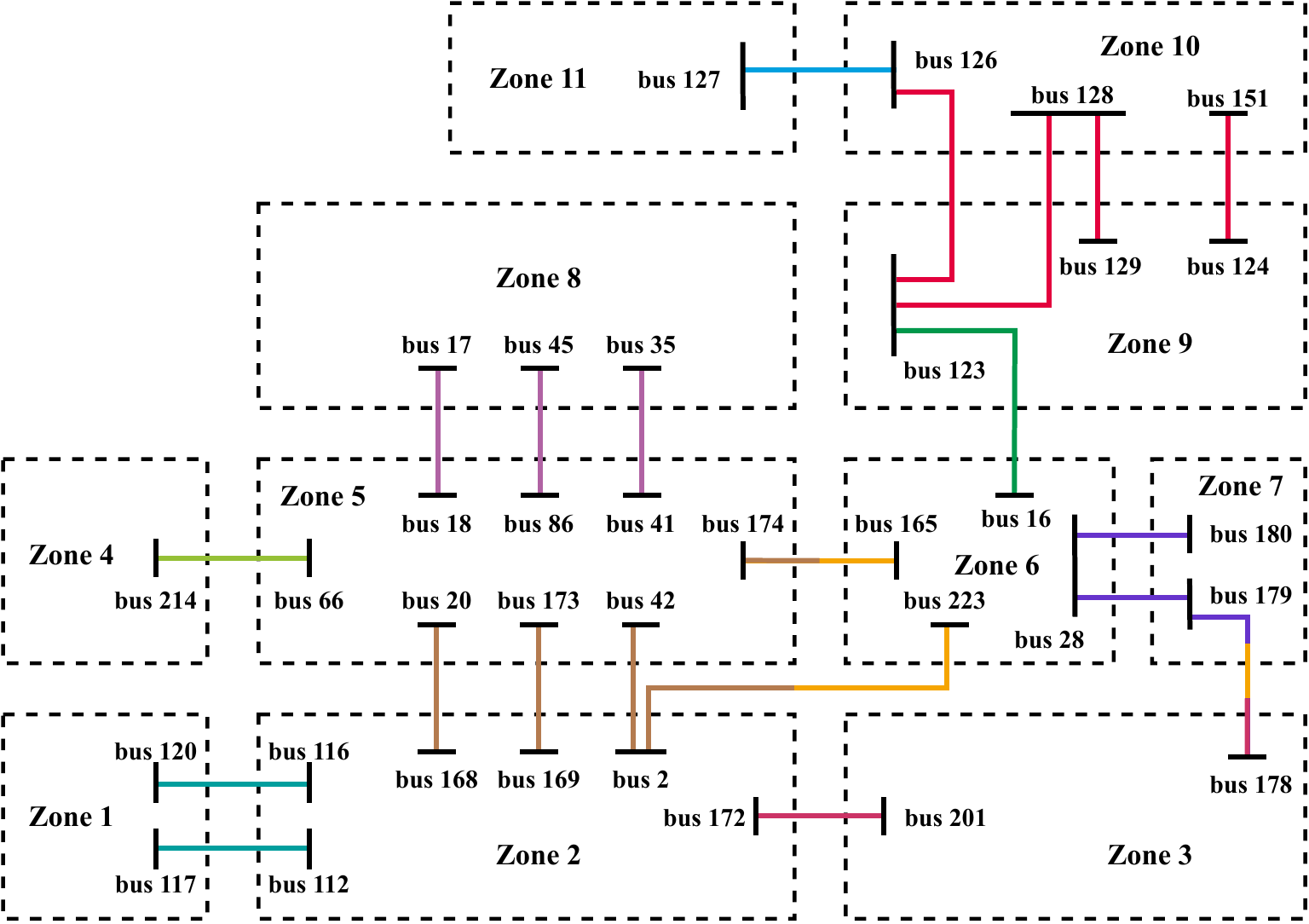}
  \caption{Illustration of the realistic 300-bus system in China. Different transmission interfaces are represented by different colors.}
  \label{fig:case300}
\end{figure}

\begin{figure}[!b]
  \centering
  \includegraphics[width=0.48\textwidth]{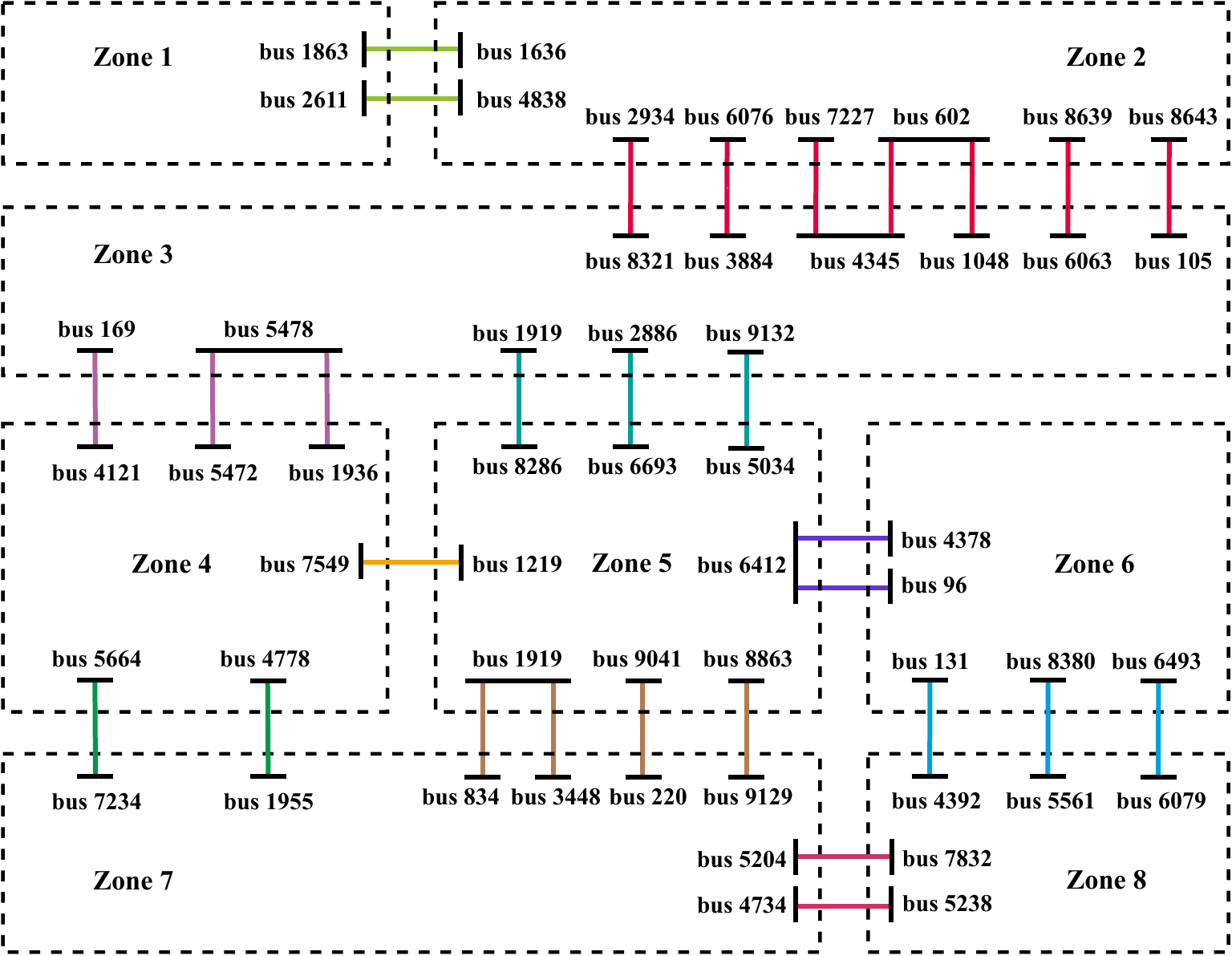}
  \caption{Illustration of the European 9241-bus system from the PEGASE project. Different transmission interfaces are represented by different colors.}
  \label{fig:case9241}
\end{figure}

\subsection{Scenario Generation}

To obtain adequate scenarios for training and testing, we first use two small-scale systems for simulations, including the IEEE 118-bus system and a realistic regional 300-bus system in China with 182 loads, 23 generators and 313 AC lines. Then we adopt a very large European 9241-bus system from the PEGASE project~\cite{fliscounakis2013contingency,josz2016ac}.
The detailed scenario generation process is summarized as follows:

\begin{itemize}[leftmargin=*]
  \item To simulate various scenarios in a given power system, we randomly select $25\%$ of the total loads and generators in each scenario, and then randomly perturb load and generator power output from $10\%$ to $200\%$ with an interval of $10\%$ of original power flow.  
  \item For different power flow adjustment tasks as shown in Figure~\ref{fig:case118}, Figure~\ref{fig:case300} and Figure~\ref{fig:case9241}, $10$ transmission interfaces are selected for each power system, respectively.
  \item The pre-scheduled power range $[\sigma_\Phi^{-}, \sigma_\Phi^+]$ of each transmission interface $\Phi$ is provided for power flow adjustment, as shown in Table~\ref{tab:range}. For each power flow adjustment task, only the insecure scenarios, of which the power flow of the corresponding transmission interfaces falls outside the pre-scheduled power range, are considered for case studies. We form a control horizon based on the dispatch action sequence. The total numbers of the insecure scenarios for each transmission interface in different systems are also listed in Table~\ref{tab:range}.
\end{itemize}

Finally, a total number of 1829 scenarios for the IEEE 118-bus system are obtained, including 1656 scenarios for training and 173 scenarios for testing. For the realistic 300-bus system in China, a total number of 1817 scenarios are obtained, including 1637 scenarios for training and 180 scenarios for testing. For the European 9241-bus system, a total number of 2037 scenarios are obtained, including 1848 scenarios for training and 189 scenarios for testing.

\begin{table}[!t]
  \centering
  \caption{The pre-scheduled power range~(MW) of each transmission interface in the IEEE 118-bus system, the realistic 300-bus system and the European 9241-bus System. ``\#'' denotes the total numbers of the insecure scenarios.}
  \label{tab:range}
  \resizebox{0.48\textwidth}{!}{%
  \begin{tabular}{@{}ccccccc@{}}
  \toprule
  \multicolumn{1}{c}{\multirow{2}{*}{\begin{tabular}[c]{@{}c@{}}\textbf{Transmission} \\ \textbf{Interface}\end{tabular}}} & \multicolumn{2}{c}{\textbf{118-bus System}} & \multicolumn{2}{c}{\textbf{300-bus System}} & \multicolumn{2}{c}{\textbf{9241-bus System}} \\ \cmidrule(l){2-7} 
  \multicolumn{1}{c}{} & \multicolumn{1}{c}{\textbf{Range}} & \multicolumn{1}{c|}{\textbf{\#}} & \multicolumn{1}{c}{\textbf{Range}} & \multicolumn{1}{c}{\textbf{\#}} & \multicolumn{1}{c}{\textbf{Range}} & \multicolumn{1}{c}{\textbf{\#}} \\ \midrule
  \textcolor[RGB]{149,192,54}{$\blacksquare$} $\Phi_1$   & $[90, 640]$  & $150$ & $[140, 1000]$  & $182$ & $[15, 110]$  & $202$         \\ \specialrule{0em}{1pt}{1pt}
  \textcolor[RGB]{227,0,61}{$\blacksquare$} $\Phi_2$     & $[50, 360]$  & $185$ & $[280, 1960]$  & $162$ & $[840, 5880]$  & $210$         \\ \specialrule{0em}{1pt}{1pt}
  \textcolor[RGB]{176,98,163}{$\blacksquare$} $\Phi_3$   & $[40, 290]$  & $211$ & $[170, 1200]$  & $174$ & $[145, 1000]$  & $204$         \\ \specialrule{0em}{1pt}{1pt}
  \textcolor[RGB]{245,165,0}{$\blacksquare$} $\Phi_4$    & $[90, 640]$  & $238$ & $[240, 1680]$  & $188$ & $[55, 390]$  & $204$         \\ \specialrule{0em}{1pt}{1pt}
  \textcolor[RGB]{0,158,157}{$\blacksquare$} $\Phi_5$    & $[70, 480]$  & $161$ & $[460, 3200]$  & $198$ & $[560, 3920]$  & $200$         \\ \specialrule{0em}{1pt}{1pt}
  \textcolor[RGB]{0,152,75}{$\blacksquare$} $\Phi_6$     & $[45, 300]$  & $185$ & $[200, 1400]$  & $204$ & $[360, 2520]$  & $203$         \\ \specialrule{0em}{1pt}{1pt}
  \textcolor[RGB]{102,51,204}{$\blacksquare$} $\Phi_7$   & $[130, 880]$  & $160$ & $[200, 1400]$  & $98$ & $[345, 2410]$  & $208$         \\ \specialrule{0em}{1pt}{1pt}
  \textcolor[RGB]{180,123,79}{$\blacksquare$} $\Phi_8$   & $[55, 390]$  & $184$ & $[480, 3360]$  & $171$ & $[760, 5320]$  & $205$         \\ \specialrule{0em}{1pt}{1pt}
  \textcolor[RGB]{204,51,102}{$\blacksquare$} $\Phi_9$   & $[130, 880]$  & $158$ & $[170, 1200]$  & $220$ & $[130, 910]$  & $200$         \\ \specialrule{0em}{1pt}{1pt}
  \,\,\,\textcolor[RGB]{0,160,223}{$\blacksquare$} $\Phi_{10}$   & $[90, 615]$  & $197$                       & $[80, 590]$  & $220$ & $[180, 1260]$  & $201$         \\  \bottomrule
  \end{tabular}%
  }
  \end{table}

\subsection{Compared Methods and Parameter Settings}

The proposed \emph{Multi-task Attribution Map} method, termed as MAM, is compared with the existing state-of-the-art DRL methods in discrete action space, including \emph{Deep Q Network}~(DQN)~\cite{DQN15}, \emph{Double DQN}~\cite{DoubleDQN}, \emph{Dueling DQN}~\cite{DuelingDQN}, \emph{Advantage Actor Critic}~(A2C)~\cite{A3C} and \emph{Proximal Policy Optimization}~(PPO)~\cite{PPO}. Two basic multi-task architectures are considered for all DRL baselines: (1) \textbf{Concat-based architecture}. Using one MLP network that takes the concat vector of the task vector and node feature vector as input. (2) \textbf{Soft-based architecture}. Using the soft modularization network~\cite{RLSoft} that enables the agent to select different MLP network modules based on the task representation.
Moreover, we also adopt \emph{Optimal Power Flow}~(OPF) method as a conventional model-based baseline based on PYPOWER~\cite{MATPOWER}. The internal solver of OPF uses the interior point method where the optimization is initialized with a valid power flow solution. All the reported results of different methods are tested on the unseen test scenarios. Three evaluation metrics are used, including test success rate, test economic cost and inference speed.

We adopt the Tianshou RL framework~\cite{tianshou} to run all experiments, which uses the PyTorch library for implementation. Moreover, power flow calculation is performed based on the PandaPower simulation~\cite{pandapower}. The detailed hyperparameters are given as follows, where the common hyperparameters across different baselines are consistent to ensure comparability and their special hyperparameters can be referred in the source code. In the proposed MAM, each GCN branch contains a $2$-layer GCN with the dimension of $[64, 64]$. The task encoder is implemented by a $2$-layer MLP network with the dimension of $[128, 64]$. For Dueling architecture, a $2$-layer value network with the dimension of $128$ and a $1$-layer advantage network with the dimension of $128$ are applied. All network modules use $\operatorname{ReLU}$ activation function. Batches of 64 episodes are sampled from the replay buffer with the size of 20K for every training iteration. The target update interval is set to 100, and the discount factor is set to 0.9. We use the Adam optimizer with a learning rate of 0.001. For exploration, $\epsilon$-greedy is used with $\epsilon$ annealed linearly from 0.1 to 0.01 over 500K training steps and kept constant for the rest of the training. Case studies are carried out on a computing platform with Intel(R) Xeon(R) Platinum 8260L CPU @ 2.40GHz, 128 GB RAM, and Quadro~P5000~GPU.

\subsection{Single-Interface Task}

\begin{figure*}[!t]
  \centering
  \subfloat{\includegraphics[scale=0.372]{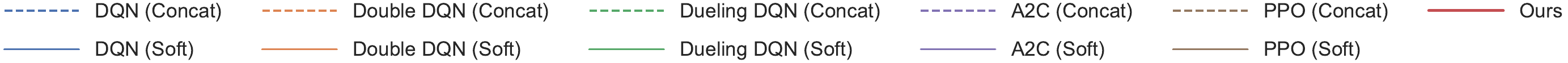}}
  \vspace{-0.1cm}
  \\
  \addtocounter{subfigure}{-1}
  \captionsetup[subfloat]{justification=centering}
  \subfloat[IEEE 118-bus System\\(Single-Interface Tasks)]{\includegraphics[width=0.3\textwidth]{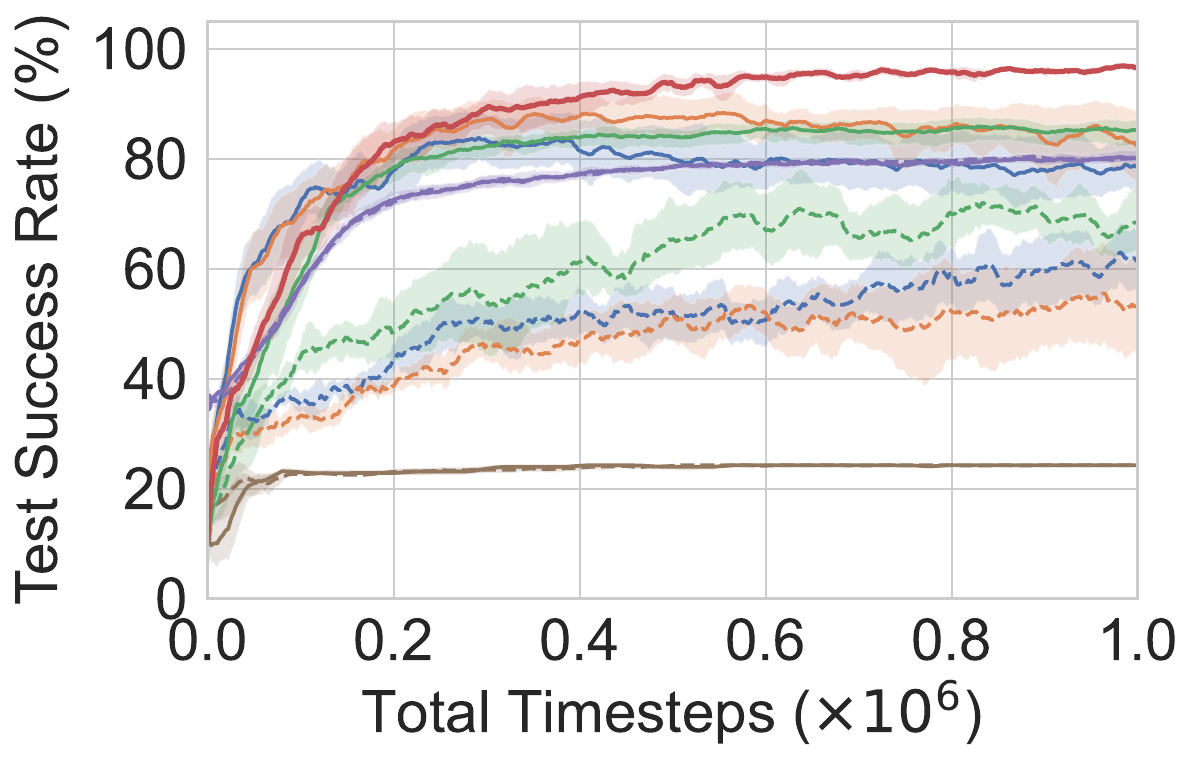}\label{fig:single10-118}} \hfil
  \subfloat[Realistic 300-bus System\\(Single-Interface Tasks)]{\includegraphics[width=0.3\textwidth]{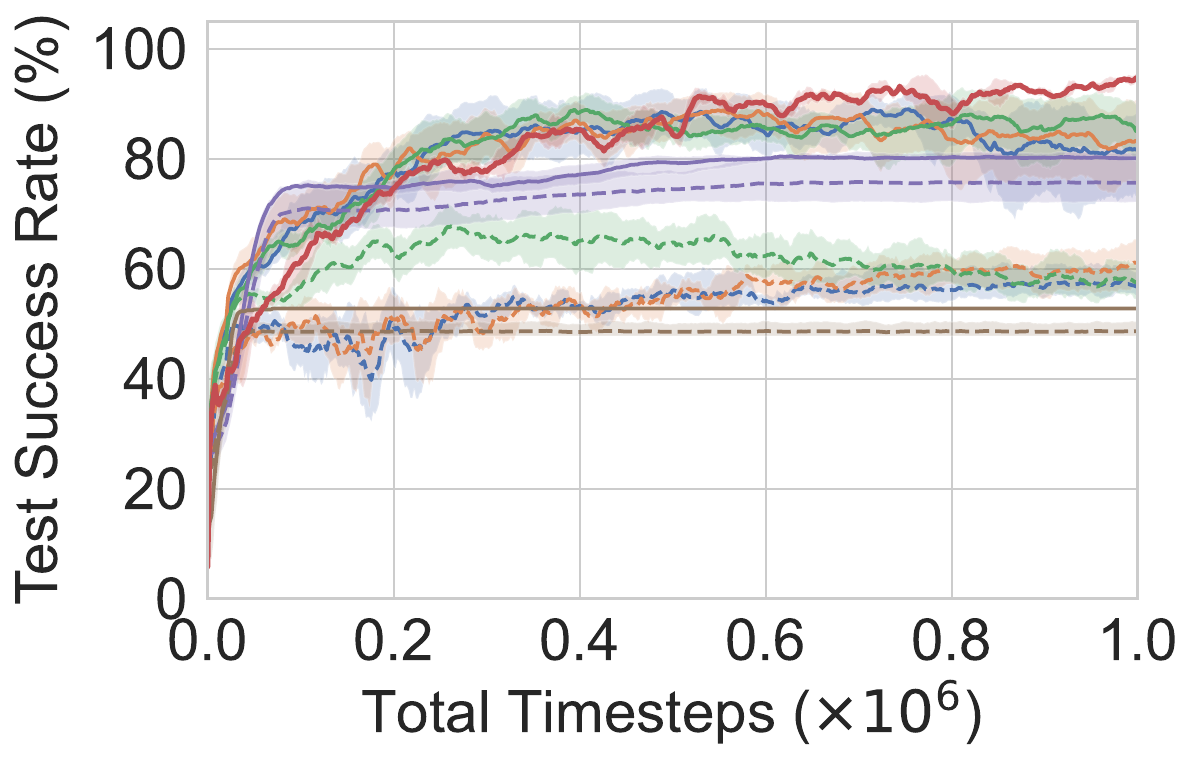}\label{fig:single10-300}} \hfil
  \subfloat[European 9241-bus System\\(Single-Interface Tasks)]{\includegraphics[width=0.3\textwidth]{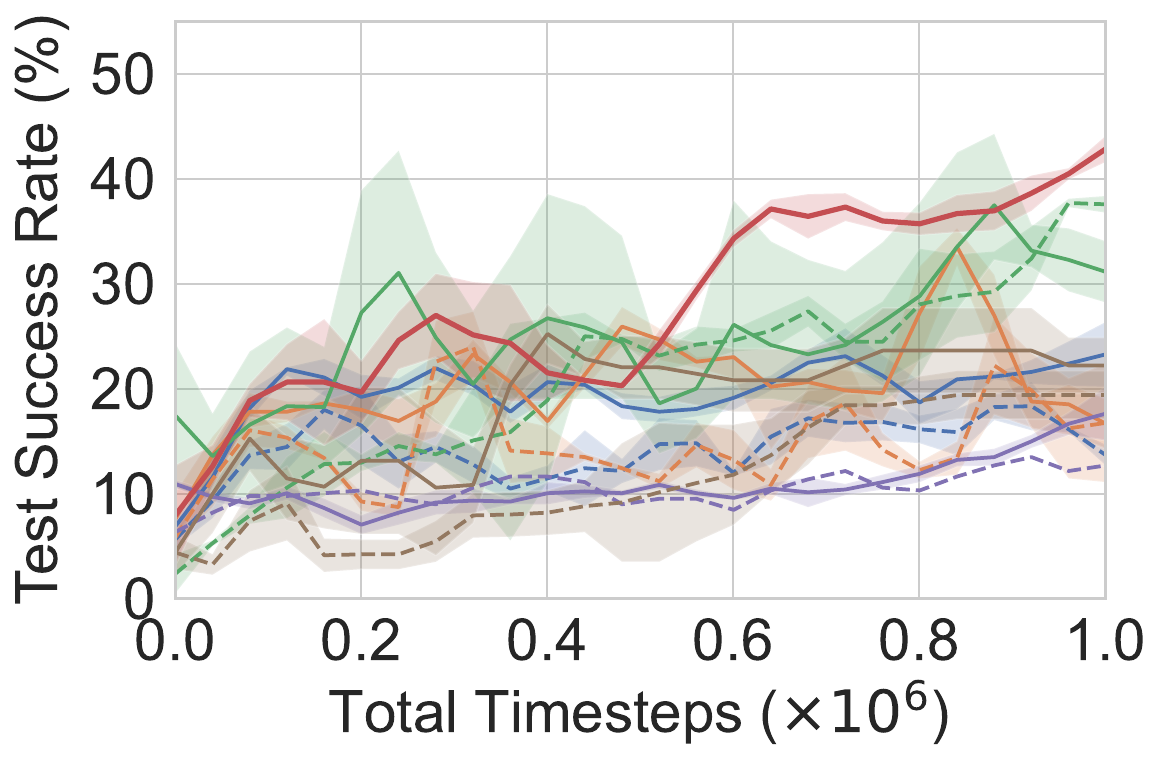}\label{fig:single10-9241}} \\
  \subfloat[IEEE 118-bus System\\(Multi-Interface Tasks)]{\includegraphics[width=0.3\textwidth]{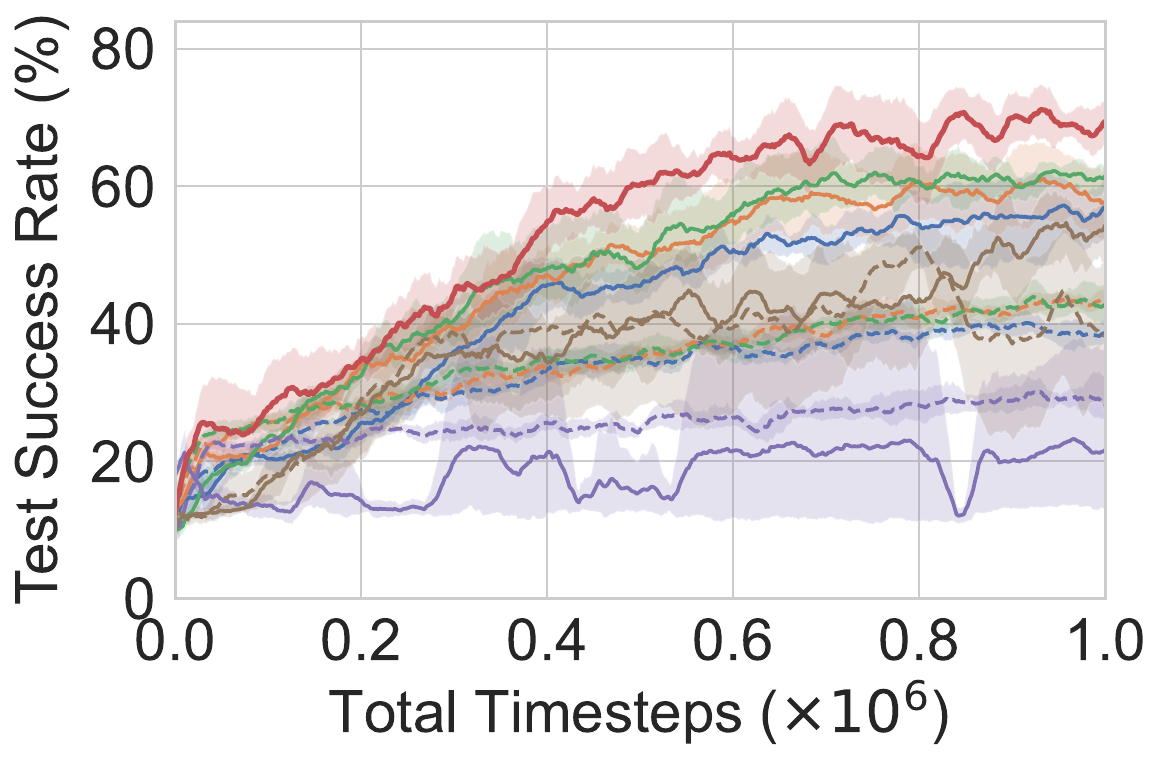}\label{fig:multi-118}} \hfil
  \subfloat[Realistic 300-bus System\\(Multi-Interface Tasks)]{\includegraphics[width=0.3\textwidth]{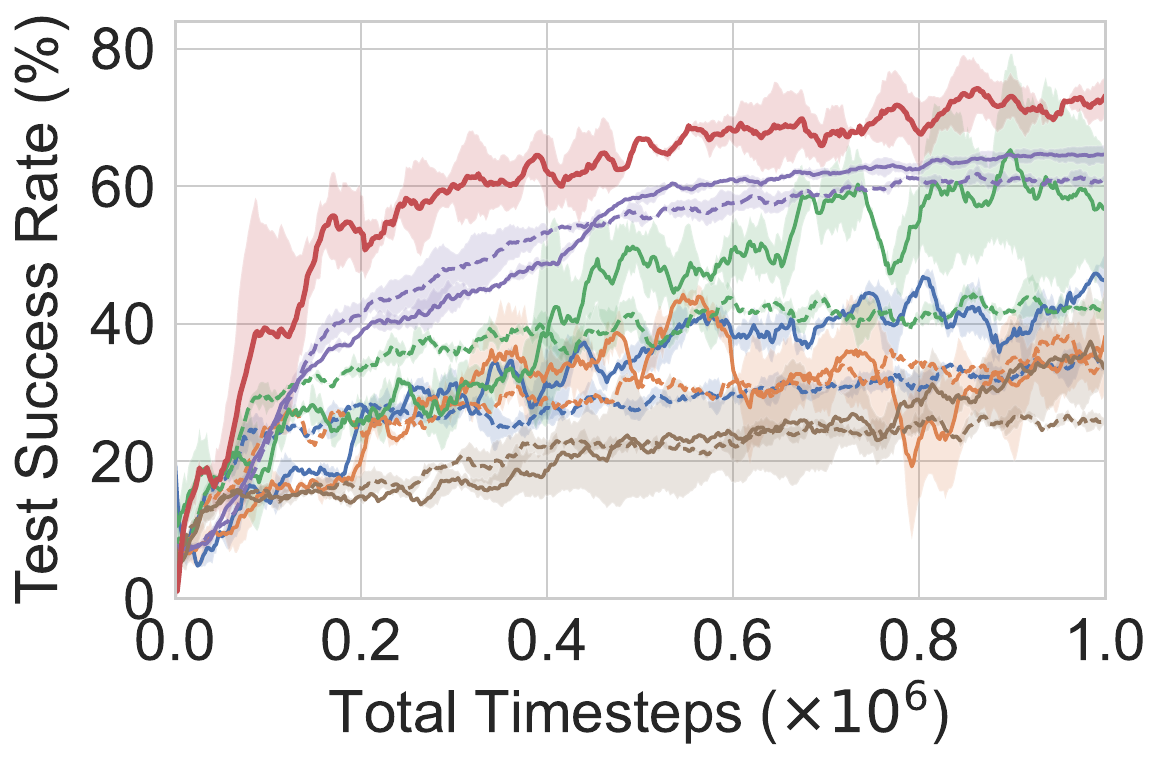}\label{fig:multi-300}} \hfil
  \subfloat[European 9241-bus System\\(Multi-Interface Tasks)]{\includegraphics[width=0.3\textwidth]{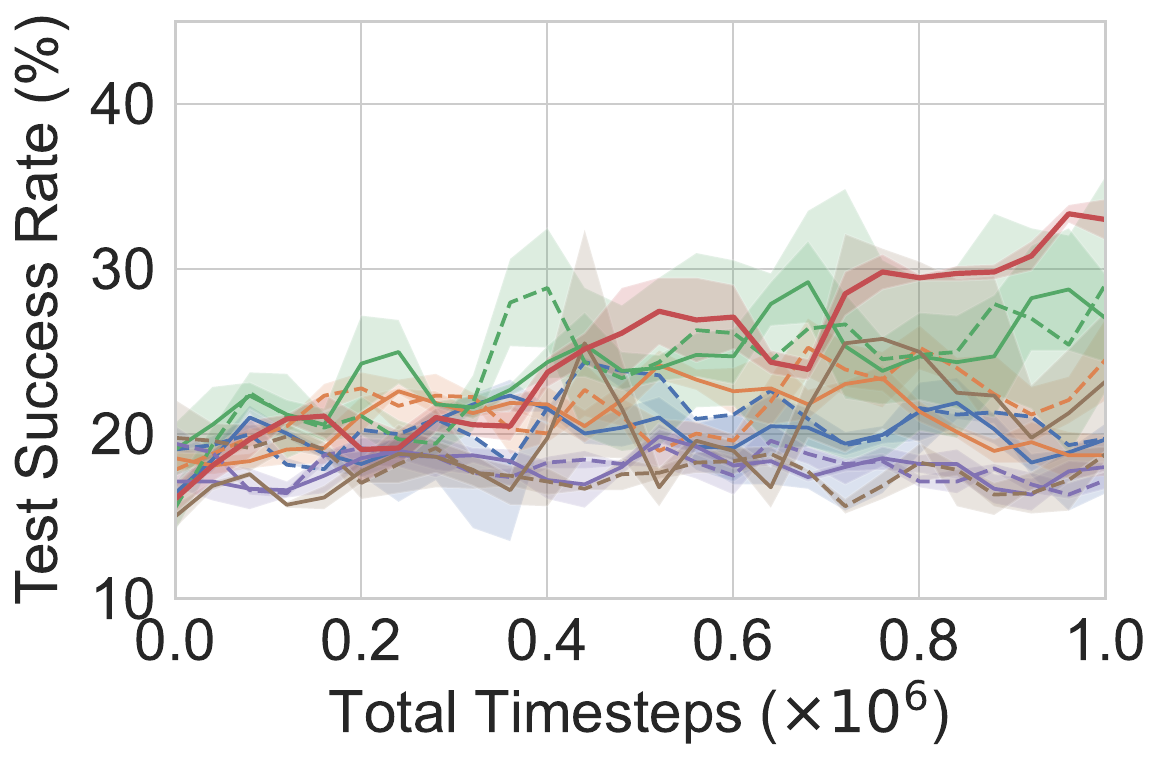}\label{fig:multi-9241}}
  \caption{Learning curves of our method and baselines in both single-interface tasks and multi-interface tasks under the multi-task setting. All experimental results are illustrated with the median performance and one standard deviation~(shaded region) over 5 random seeds for a fair comparison.}
  \label{fig:single}
\end{figure*}

To validate the effectiveness of our method in single-interface tasks under the multi-task setting, we conduct experiments on both small-scale and large-scale power systems. For each power system, 10 selected transmission interfaces are considered as 10 single-interface tasks. We consider the multi-task setting with all single-interface tasks $\left\{\mathcal{T}(\Phi_1),\mathcal{T}(\Phi_2),\cdots,\mathcal{T}(\Phi_{10})\right\}$. Each single-interface task is randomly sampled at each episode. The learning curves of different DRL methods on different power system are shown in Figure~\ref{fig:single10-118}--\ref{fig:single10-9241}. The final performance of all DRL and OPF methods is shown in Table~\ref{tab:single}. The average inference speed of our method and OPF baselines is shown in Table~\ref{tab:single-time}.

\begin{table}[!t]
  \centering
  \caption{The performance of our method and baselines in single-interface tasks under the multi-task setting. The performance is obtained under the average evaluation over 5 trials. The performance differences between our method and baselines are given in parentheses. Note that the higher test success rate and the lower test economic cost, the better performance of the method.}
  \label{tab:single}
  \resizebox{0.48\textwidth}{!}{%
  \begin{tabular}{@{}lccc@{}}
    \toprule
    \multicolumn{1}{c}{\multirow{2}{*}{\textbf{Method}}} & \multicolumn{3}{c}{\textbf{Test Success Rate~(\%)}} \\ \cmidrule(l){2-4} 
    & \multicolumn{1}{c}{\textbf{118-bus System}} & \multicolumn{1}{c}{\textbf{300-bus System}} & \multicolumn{1}{c}{\textbf{9241-bus System}} \\ \midrule
    DQN (Concat)               & 61.23{\tiny{~(-35.38)}} & 56.74{\tiny{~(-38.06)}} & 13.67{\tiny{~(-29.19)}}            \\  \specialrule{0em}{1pt}{1pt}
    DQN (Soft)                 & 78.53{\tiny{~(-18.08)}} & 81.88{\tiny{~(-12.92)}} & 23.28{\tiny{~(-19.58)}}            \\ \specialrule{0em}{1pt}{1pt}
    Double DQN (Concat)        & 53.28{\tiny{~(-43.33)}} & 61.24{\tiny{~(-33.56)}} & 16.75{\tiny{~(-26.11)}}             \\ \specialrule{0em}{1pt}{1pt}
    Double DQN (Soft)          & 82.35{\tiny{~(-14.26)}} & 83.11{\tiny{~(-11.69)}} & 16.67{\tiny{~(-26.17)}}            \\ \specialrule{0em}{1pt}{1pt}
    Dueling DQN (Concat)       & 68.44{\tiny{~(-28.17)}} & 57.66{\tiny{~(-37.14)}} & 37.57{\;\tiny{~(-5.29)}}            \\ \specialrule{0em}{1pt}{1pt}
    Dueling DQN (Soft)         & 85.23{\tiny{~(-11.38)}} & 85.30{\;\tiny{~(-9.50)}} & 31.13{\tiny{~(-11.73)}}           \\ \specialrule{0em}{1pt}{1pt}
    A2C (Concat)               & 80.14{\tiny{~(-16.47)}} & 75.61{\tiny{~(-19.19)}} & 12.70{\tiny{~(-30.16)}}            \\ \specialrule{0em}{1pt}{1pt}
    A2C (Soft)                 & 80.10{\tiny{~(-16.51)}} & 80.14{\tiny{~(-14.66)}} & 17.64{\tiny{~(-25.22)}}            \\ \specialrule{0em}{1pt}{1pt}
    PPO (Concat)               & 24.28{\tiny{~(-72.33)}} & 48.61{\tiny{~(-46.19)}} & 19.40{\tiny{~(-23.46)}}            \\ \specialrule{0em}{1pt}{1pt}
    PPO (Soft)                 & 24.28{\tiny{~(-72.33)}} & 52.78{\tiny{~(-42.02)}} & 22.22{\tiny{~(-20.64)}}            \\ \specialrule{0em}{1pt}{1pt}
    OPF                 & 72.25{\tiny{~(-24.36)}} & 50.56{\tiny{~(-44.24)}} & 51.32{\;\tiny{~(+8.46)}}            \\ \midrule
    \textbf{MAM~(Ours)}        & \textbf{96.61} & \textbf{94.80} & \textbf{42.86}           \\ \specialrule{0em}{1pt}{1pt} \toprule
    \multicolumn{1}{c}{\multirow{2}{*}{\textbf{Method}}} & \multicolumn{3}{c}{\textbf{Test Economic Cost~(\$)}} \\ \cmidrule(l){2-4}
    & \multicolumn{1}{c}{\textbf{118-bus System}} & \multicolumn{1}{c}{\textbf{300-bus System}} & \multicolumn{1}{c}{\textbf{9241-bus System}} \\ \midrule
    Original               & 632,354{\;\tiny{~(+8,901)}} & 999,927{\tiny{~(+29,669)}} & 325,606{\tiny{~(+1,112)}}           \\  \specialrule{0em}{1pt}{1pt}
    DQN (Concat)           & 626,808{\;\tiny{~(+3,355)}} & 991,762{\tiny{~(+21,504)}} & 324,658{\tiny{\;\;\;~(+164)}}           \\  \specialrule{0em}{1pt}{1pt}
    DQN (Soft)                 & 625,598{\;\tiny{~(+2,145)}} & 986,337{\tiny{~(+16,079)}} & 325,554{\tiny{~(+1,060)}}           \\ \specialrule{0em}{1pt}{1pt}
    Double DQN (Concat)        & 638,079{\tiny{~(+14,626)}} & 990,126{\tiny{~(+19,868)}} & 324,792{\tiny{\;\;\;~(+298)}}           \\ \specialrule{0em}{1pt}{1pt}
    Double DQN (Soft)          & 625,279{\;\tiny{~(+1,826)}} & 982,362{\tiny{~(+12,104)}} & 325,790{\tiny{~(+1,296)}}            \\ \specialrule{0em}{1pt}{1pt}
    Dueling DQN (Concat)       & 626,215{\;\tiny{~(+2,762)}} & 992,048{\tiny{~(+21,790)}} & 324,485{\tiny{\;\;\;\;\;\;\;~(-9)}}            \\ \specialrule{0em}{1pt}{1pt}
    Dueling DQN (Soft)         & 625,128{\;\tiny{~(+1,678)}} & 972,573{\;\tiny{~(+2,315)}} & 325,152{\tiny{\;\;\;~(+658)}}            \\ \specialrule{0em}{1pt}{1pt}
    A2C (Concat)               & 626,703{\;\tiny{~(+3,250)}} & 976,467{\;\tiny{~(+6,209)}} & 325,000{\tiny{\;\;\;~(+506)}}           \\ \specialrule{0em}{1pt}{1pt}
    A2C (Soft)                 & 625,381{\;\tiny{~(+1,928)}} & 976,803{\;\tiny{~(+6,545)}} & 324,808{\tiny{\;\;\;~(+314)}}           \\ \specialrule{0em}{1pt}{1pt}
    PPO (Concat)               & 596,315{\tiny{~(-27,138)}} & 930,581{\tiny{~(-39,677)}} & 325,549{\tiny{~(+1,055)}}          \\ \specialrule{0em}{1pt}{1pt}
    PPO (Soft)                 & 596,577{\tiny{~(-26,876)}} & 936,181{\tiny{~(-34,077)}} & 325,303{\tiny{\;\;\;~(+809)}}           \\ \specialrule{0em}{1pt}{1pt}
    OPF                        & 606,259{\tiny{~(-17,194)}} & 953,349{\tiny{~(-16,909)}} & 321,586{\tiny{~(-2,908)}}              \\ \midrule
    \textbf{MAM~(Ours)}              & \textbf{623,453}        & \textbf{970,258}   & \textbf{324,494}          \\ \bottomrule
    \end{tabular}%
    }
\end{table}

\begin{table}[!t]
  \centering
  \caption{The average inference speed of our method and OPF baselines in single-interface tasks. $\pm$ corresponds to one standard deviation of the average evaluation over all test scenarios.}
  \label{tab:single-time}
  \resizebox{0.48\textwidth}{!}{%
  \begin{tabular}{@{}lccc@{}}
    \toprule
    \multicolumn{1}{c}{\multirow{2}{*}{\textbf{Method}}} & \multicolumn{3}{c}{\textbf{Inference Time~(s)}} \\ \cmidrule(l){2-4}
    & \multicolumn{1}{c}{\textbf{118-bus System}} & \multicolumn{1}{c}{\textbf{300-bus System}} & \multicolumn{1}{c}{\textbf{9241-bus System}}\\ \midrule
    OPF                        & 44.822 $\pm$ 38.459       & 47.668 $\pm$   28.230   & 1912.608 $\pm$   1636.746        \\ \specialrule{0em}{1pt}{1pt}
    \textbf{MAM~(Ours)}              & \textbf{0.006 $\pm$  0.009}      & \textbf{0.010 $\pm$  0.011}    & \textbf{0.051 $\pm$  0.025}        \\  \bottomrule
    \end{tabular}%
    }
\end{table}

In the IEEE 118-bus system and the realistic 300-bus system, our proposed MAM successfully improves the final performance and offers a very high inference speed. (1)~From the perspective of different multi-task architectures, the soft-based architecture consistently outperforms the concat-based one, while our proposed method can still achieve the best performance. The idea of the soft-based architecture is similar to ours. It focuses on learning the relationship between network modules and tasks, but ignores the relationship between input nodes and tasks. Different node features are still coupled together, leading to poorly generalizing at test time due to over-fitting on noisy nodes during the training time. Thus, the proposed MAM takes advantage of the attribution map to selectively integrate the node features into a compact state representation for the final policy. (2)~Comparing different DRL methods, despite DQN and its variants showing high learning efficiency, these methods inevitably fall into suboptimal policies with low test success rates. The performance of the A2C algorithm is worse than Dueling DQN, while PPO cannot learn any effective policy and perform poorly. In contrast, our proposed MAM works quite well on the test success rate and outperforms all state-of-the-art DRL methods during training. Furthermore, the proposed MAM not only achieves the best test success rate, but also obtains the lowest test economic cost except for PPO. Despite the lower test economic cost obtained by PPO, it is worth noting that the success rate of PPO is extraordinarily low. PPO only arbitrarily reduces the power of the generators without considering the power flow target of the transmission interface, which is unacceptable. DQN and its variants are off-policy algorithms, while A2C and PPO are on-policy algorithms. Off-policy algorithm enables the agent to update its policy based on the interaction samples from the historical policy, but on-policy algorithm only uses the interaction samples from the current policy. Therefore, the sample efficiency of off-policy algorithm is higher than the on-policy algorithm, which is beneficial in the power system environment where the control sequence is short. On the other hand, PPO provides a policy constraint over the A2C algorithm, which may lead to a locally optimal solution due to the short control sequence. 

It is notable that our MAM also achieves superior performance compared to the conventional OPF method in the small-scale system. We only focus on the insecure scenarios for each specific transmission interfaces, where the OPF method requires solving complex nonlinear optimization problems. Although the OPF method can easily obtain a lower test economic cost, it often fails to satisfy the power flow constraints and perform poorly on the test success rate. However, our MAM can learn to generalize different scenarios and enables a near-optimal decision, making it more robust and flexible. The results suggest that the attribution map can be utilized to explore the diverse control mechanisms for different tasks, which helps the agents to construct a more useful policy and achieves non-trivial performance.

In the 9241-bus system, the problems of power flow adjustment become much more challenging due to the large exploration space. We can observe that all DRL baselines cannot perform well, while only OPF and MAM can still offer the reasonable results. However, the conventional OPF method become computationally expensive as the system size increases. It requires a lot of time to find the solution in the nonlinear nonconvex searching space. On the contrary, the model-free MAM method can directly use neural network forward propagation to deliver dispatch action in a more efficient manner. Thus, MAM can provide competent inference speed guarantees for practical deployment, while the time consumption of OPF is unacceptable. Moreover, MAM also exhibits its advantage in facilitates filtering the noisy information and yields performance on par with OPF in~the~large-scale~system.

\begin{table}[!t]
  \centering
  \caption{The performance of our method and baselines in multi-interface tasks under the multi-task setting.}
  \label{tab:multi}
  \resizebox{0.48\textwidth}{!}{%
  \begin{tabular}{@{}lccc@{}}
    \toprule
    \multicolumn{1}{c}{\multirow{2}{*}{\textbf{Method}}} & \multicolumn{3}{c}{\textbf{Test Success Rate~(\%)}} \\ \cmidrule(l){2-4} 
    & \multicolumn{1}{c}{\textbf{118-bus System}} & \multicolumn{1}{c}{\textbf{300-bus System}} & \multicolumn{1}{c}{\textbf{9241-bus System}} \\ \midrule
    DQN (Concat)               & 38.66{\tiny{~(-30.79)}} & 34.12{\tiny{~(-39.06)}}  & 19.71{\tiny{~(-13.27)}}           \\  \specialrule{0em}{1pt}{1pt}
    DQN (Soft)                 & 56.92{\tiny{~(-12.53)}} & 46.26{\tiny{~(-26.92)}}  & 19.66{\tiny{~(-13.32)}}           \\ \specialrule{0em}{1pt}{1pt}
    Double DQN (Concat)        & 43.52{\tiny{~(-25.93)}} & 36.26{\tiny{~(-36.92)}}   & 24.51{\;\tiny{~(-8.47)}}          \\ \specialrule{0em}{1pt}{1pt}
    Double DQN (Soft)          & 57.44{\tiny{~(-12.01)}} & 38.26{\tiny{~(-34.92)}}   & 18.69{\tiny{~(-14.29)}}          \\ \specialrule{0em}{1pt}{1pt}
    Dueling DQN (Concat)       & 42.65{\tiny{~(-26.80)}} & 42.15{\tiny{~(-31.03)}}   & 29.10{\;\tiny{~(-3.88)}}          \\ \specialrule{0em}{1pt}{1pt}
    Dueling DQN (Soft)         & 61.54{\;\tiny{~(-7.91)}} & 56.67{\tiny{~(-16.51)}}  & 26.98{\;\tiny{~(-6.00)}}           \\ \specialrule{0em}{1pt}{1pt}
    A2C (Concat)               & 28.19{\tiny{~(-41.26)}} & 60.98{\tiny{~(-12.20)}}   & 17.20{\tiny{~(-15.78)}}          \\ \specialrule{0em}{1pt}{1pt}
    A2C (Soft)                 & 21.69{\tiny{~(-47.76)}} & 64.57{\;\tiny{~(-8.61)}}  & 17.99{\tiny{~(-14.99)}}           \\ \specialrule{0em}{1pt}{1pt}
    PPO (Concat)               & 38.25{\tiny{~(-31.20)}} & 25.68{\tiny{~(-47.50)}}   & 18.78{\tiny{~(-14.20)}}          \\ \specialrule{0em}{1pt}{1pt}
    PPO (Soft)                 & 54.37{\tiny{~(-15.08)}} & 33.41{\tiny{~(-39.77)}}   & 23.19{\;\tiny{~(-9.79)}}          \\ \specialrule{0em}{1pt}{1pt}
    OPF                        & 47.98{\tiny{~(-21.47)}}         & 13.89{\tiny{~(-59.29)}}     & 37.57{\;\tiny{~(+4.59)}}          \\ \midrule
    \textbf{MAM~(Ours)}              & \textbf{69.45}        & \textbf{73.18}   & \textbf{32.98}         \\ \specialrule{0em}{1pt}{1pt} \toprule
    \multicolumn{1}{c}{\multirow{2}{*}{\textbf{Method}}} & \multicolumn{3}{c}{\textbf{Test Economic Cost~(\$)}}  \\ \cmidrule(l){2-4} 
    & \multicolumn{1}{c}{\textbf{118-bus System}} & \multicolumn{1}{c}{\textbf{300-bus System}} & \multicolumn{1}{c}{\textbf{9241-bus System}} \\ \midrule
    Original               & 632,354{\tiny{~(+57,392)}}        & 999,927{\tiny{~(+87,212)}}    & 325,606{\tiny{~(+1,180)}}        \\  \specialrule{0em}{1pt}{1pt}
    DQN (Concat)               & 619,440{\tiny{~(+44,478)}}        & 945,438{\tiny{~(+32,723)}}  & 325,316{\;\;\tiny{~(+890)}}          \\  \specialrule{0em}{1pt}{1pt}
    DQN (Soft)                 & 577,665{\;\tiny{~(+2,703)}}        & 941,011{\tiny{~(+28,296)}}  & 325,662{\tiny{~(+1,236)}}          \\ \specialrule{0em}{1pt}{1pt}
    Double DQN (Concat)        & 597,228{\tiny{~(+22,266)}}        & 943,314{\tiny{~(+30,599)}}   & 327,074{\tiny{~(+2,648)}}         \\ \specialrule{0em}{1pt}{1pt}
    Double DQN (Soft)          & 580,558{\;\tiny{~(+5,596)}}        & 955,407{\tiny{~(+42,692)}}  & 326,951{\tiny{~(+2,525)}}          \\ \specialrule{0em}{1pt}{1pt}
    Dueling DQN (Concat)       & 582,386{\;\tiny{~(+7,424)}}        & 924,819{\tiny{~(+12,104)}}  & 324,533{\;\;\tiny{~(+107)}}          \\ \specialrule{0em}{1pt}{1pt}
    Dueling DQN (Soft)         & 576,127{\;\tiny{~(+1,165)}}        & 920,950{\;\tiny{~(+8,235)}}  & 324,855{\;\;\tiny{~(+429)}}          \\ \specialrule{0em}{1pt}{1pt}
    A2C (Concat)               & 606,632{\tiny{~(+31,670)}}        & 921,492{\;\tiny{~(+8,777)}}  & 325,626{\tiny{~(+1,200)}}          \\ \specialrule{0em}{1pt}{1pt}
    A2C (Soft)                 & 625,292{\tiny{~(+50,330)}}        & 916,958{\;\tiny{~(+4,243)}}  & 325,659{\tiny{~(+1,233)}}          \\ \specialrule{0em}{1pt}{1pt}
    PPO (Concat)               & 599,601{\tiny{~(+24,639)}}        & 938,474{\tiny{~(+25,759)}}   & 325,456{\tiny{~(+1,030)}}         \\ \specialrule{0em}{1pt}{1pt}
    PPO (Soft)                 & 605,275{\tiny{~(+30,313)}}        & 936,014{\tiny{~(+23,299)}}   & 325,547{\tiny{~(+1,121)}}         \\ \specialrule{0em}{1pt}{1pt}
    OPF                        & 554,691{\tiny{~(-20,271)}}        & 956,624{\tiny{~(+43,909)}}   & 321,694{\tiny{~(-2,732)}}           \\ \midrule 
    \textbf{MAM~(Ours)}              & \textbf{574,962}        & \textbf{912,715}    & \textbf{324,426}          \\ \bottomrule
    \end{tabular}%
    }
\end{table}

\begin{table}[!t]
  \centering
  \caption{The average inference speed of our method and OPF baselines in multi-interface tasks.}
  \label{tab:multi-time}
  \resizebox{0.48\textwidth}{!}{%
  \begin{tabular}{@{}lccc@{}}
    \toprule
    \multicolumn{1}{c}{\multirow{2}{*}{\textbf{Method}}} & \multicolumn{3}{c}{\textbf{Inference Time~(s)}} \\ \cmidrule(l){2-4}
    & \multicolumn{1}{c}{\textbf{118-bus System}} & \multicolumn{1}{c}{\textbf{300-bus System}} & \multicolumn{1}{c}{\textbf{9241-bus System}}\\ \midrule
    OPF                        & 67.958 $\pm$ 30.218      & 57.729 $\pm$  23.794    & 2416.349 $\pm$  2016.855         \\ \specialrule{0em}{1pt}{1pt}
    \textbf{MAM~(Ours)}              & \textbf{0.011  $\pm$ 0.003}    & \textbf{0.030   $\pm$   0.032}  & \textbf{0.066   $\pm$   0.016}     \\ \bottomrule
    \end{tabular}%
    }
\end{table}

\subsection{Multi-Interface Task}

To further evaluate the proposed method in multi-interface tasks under the multi-task setting, we also conduct experiments on two power systems. In the IEEE 118-bus system, we consider the multi-task setting with different 5-interface tasks $\left\{\mathcal{T}(\{\Phi_m\}_{m\in\mathcal{M}})\right\}$, where $\mathcal{M}$ is the set of 5 transmission interfaces that randomly selected from all transmission interfaces. In the more complex realistic 300-bus system and European 9241-bus system, we consider the multi-task setting with different 3-interface tasks. Figure~\ref{fig:multi-118}--\ref{fig:multi-9241} present the learning curves of comparison methods on different power systems, while the final performance is shown in Table~\ref{tab:multi} and the average inference speed is shown in Table~\ref{tab:multi-time}.

In different power systems, the simulation results are similar to that of single-interface tasks. The soft-based architecture still outperforms the concat-based one from the perspective of different multi-task architectures, while our MAM approach consistently exceeds all baselines by a large margin not only in test success rate but also in test economic cost. Compared with the conventional OPF-based method, MAM significantly reduce the computational cost. The proposed MAM is particularly beneficial in these multi-interface tasks, which enables us to explore the diverse common critical nodes and generalize across various tasks. Notably, despite the encouraging results achieved, our proposed MAM cannot achieve the same high test success rate in the multi-interface tasks as in the single-interface tasks. The multi-interface tasks are more complicated than single-interface tasks because of the coupling relationship between different transmission interfaces. In some extreme operating condition scenarios, there may even be no adjustment policy that can successfully satisfy all power flow constraints. 

\begin{figure}[!t]
  \centering
  \subfloat{\includegraphics[width=0.34\textwidth]{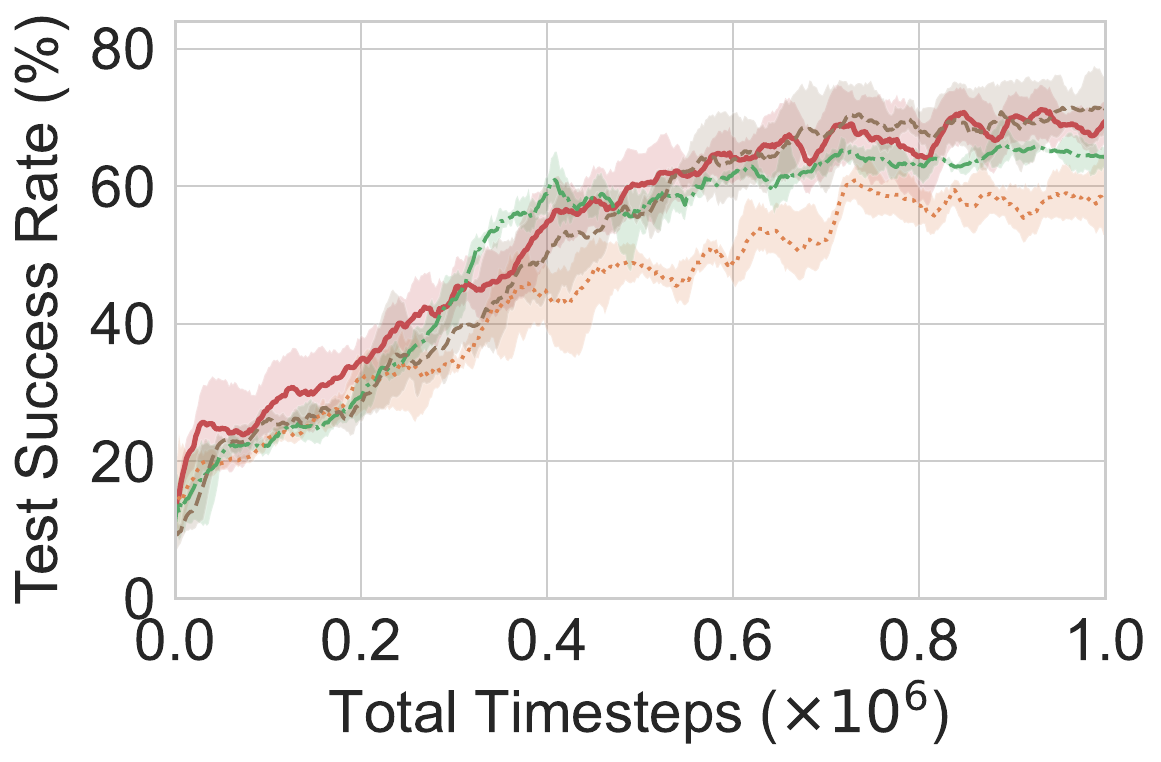}}
  \subfloat{\raisebox{0.9\height}{\includegraphics[width=0.12\textwidth]{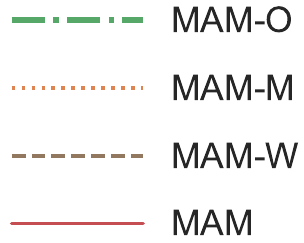}}}
  \caption{Learning curves of our proposed MAM and its ablations on the IEEE 118-bus system with multi-interface tasks.}
  \label{fig:ablation}
\end{figure}

\begin{table}[!t]
  \centering
  \caption{The performance of our proposed MAM and its ablations on the IEEE 118-bus system with multi-interface tasks.}
  \label{tab:ablation}
  \resizebox{0.48\textwidth}{!}{%
  \large
  \begin{tabular}{@{}lccc@{}}
    \toprule \specialrule{0em}{1pt}{1pt}
    \textbf{\;Method} & \textbf{Test Success Rate~(\%)}
    & \textbf{Test Economic Cost~(\$)} & \textbf{Inference Time~(s)} \\ \specialrule{0em}{1pt}{1pt} \midrule \specialrule{0em}{1pt}{1pt}
    MAM-O                        & 64.16{\;\small{~(-5.29)}}       & 581,607{\;\small{~(+6,645)}}  & 0.010{\small{~(-0.001)}}       \\ \specialrule{0em}{1pt}{1pt}
    MAM-M                        & 58.07{\small{~(-11.38)}}       & 591,698{\small{~(+16,736)}}  & 0.007{\small{~(-0.004)}}       \\ \specialrule{0em}{1pt}{1pt}
    MAM-W                        & 71.19{\;\small{~(+1.74)}}       & 602,992{\small{~(+28,030)}}   & 0.012{\small{~(+0.001)}}        \\ \specialrule{0em}{1pt}{1pt} \midrule \specialrule{0em}{1pt}{1pt}
    \textbf{MAM}                          & \textbf{69.45}      & \textbf{574,962}    & \textbf{0.011}      \\ \specialrule{0em}{1pt}{1pt}\bottomrule
    \end{tabular}%
    }
\end{table}

\begin{figure*}[!h]
  \centering
  \subfloat{\includegraphics[width=0.33\textwidth]{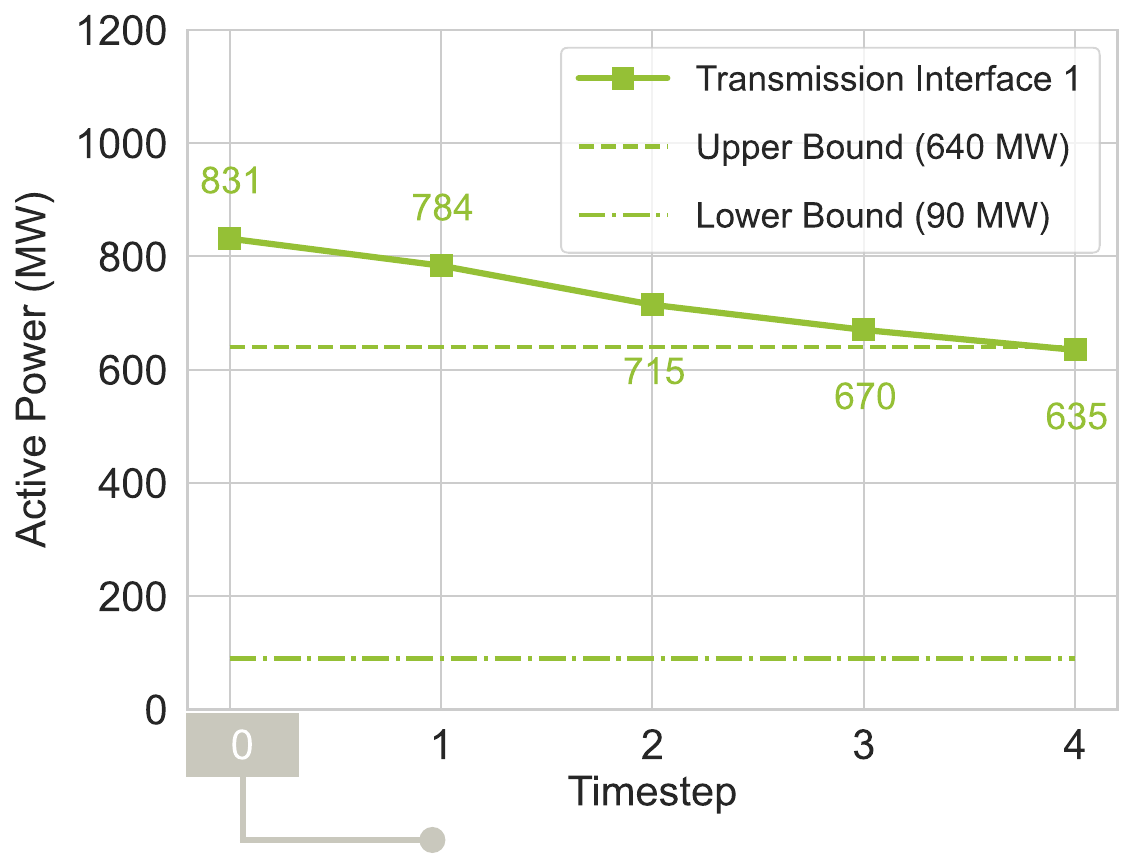}\label{fig:s2_power}}
  \subfloat{\includegraphics[width=0.33\textwidth]{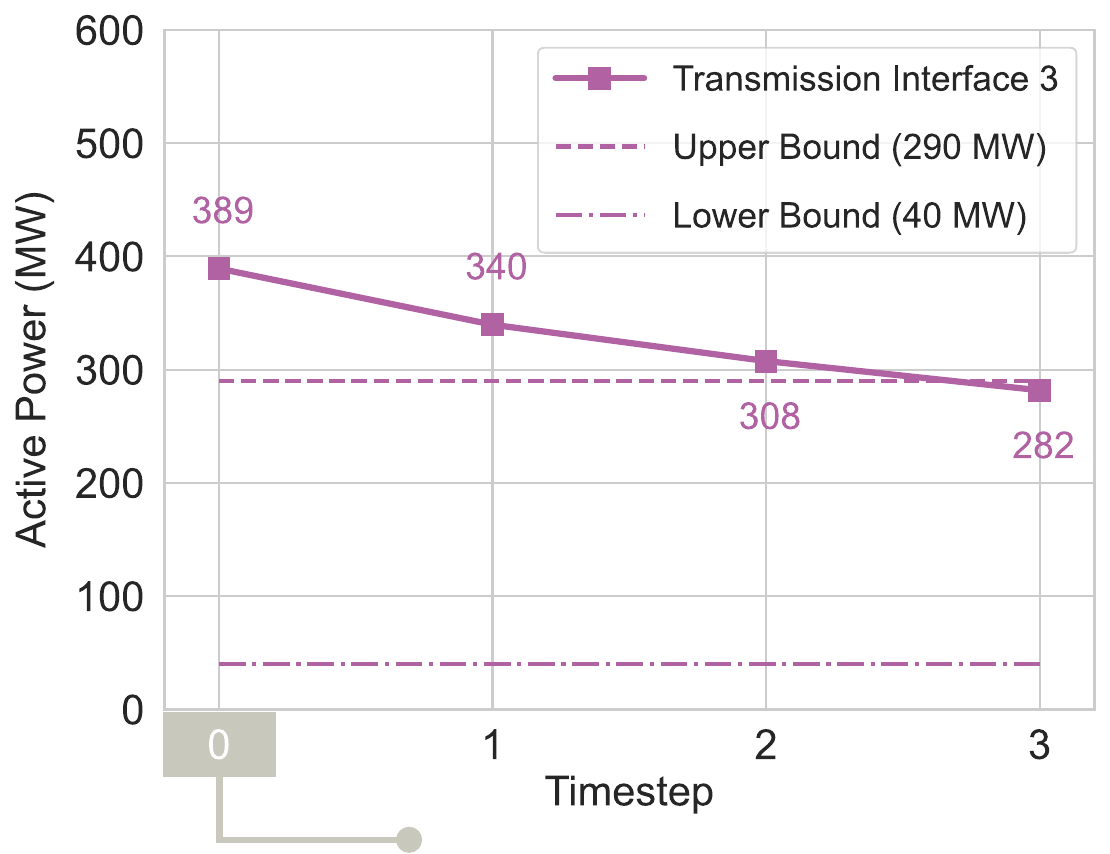}\label{fig:s3_power}}
  \subfloat{\includegraphics[width=0.33\textwidth]{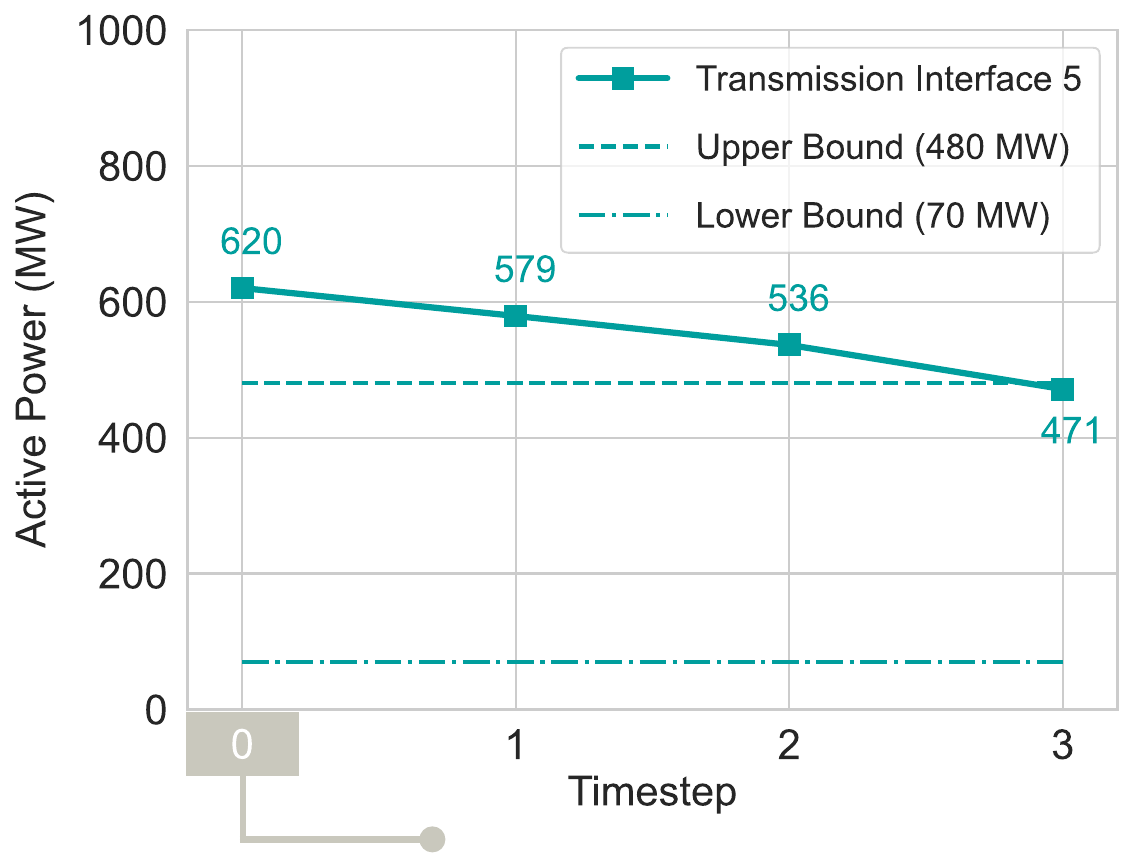}\label{fig:s1_power}}\\
  \vspace{-0.8em}
  \subfloat{\includegraphics[scale=0.22]{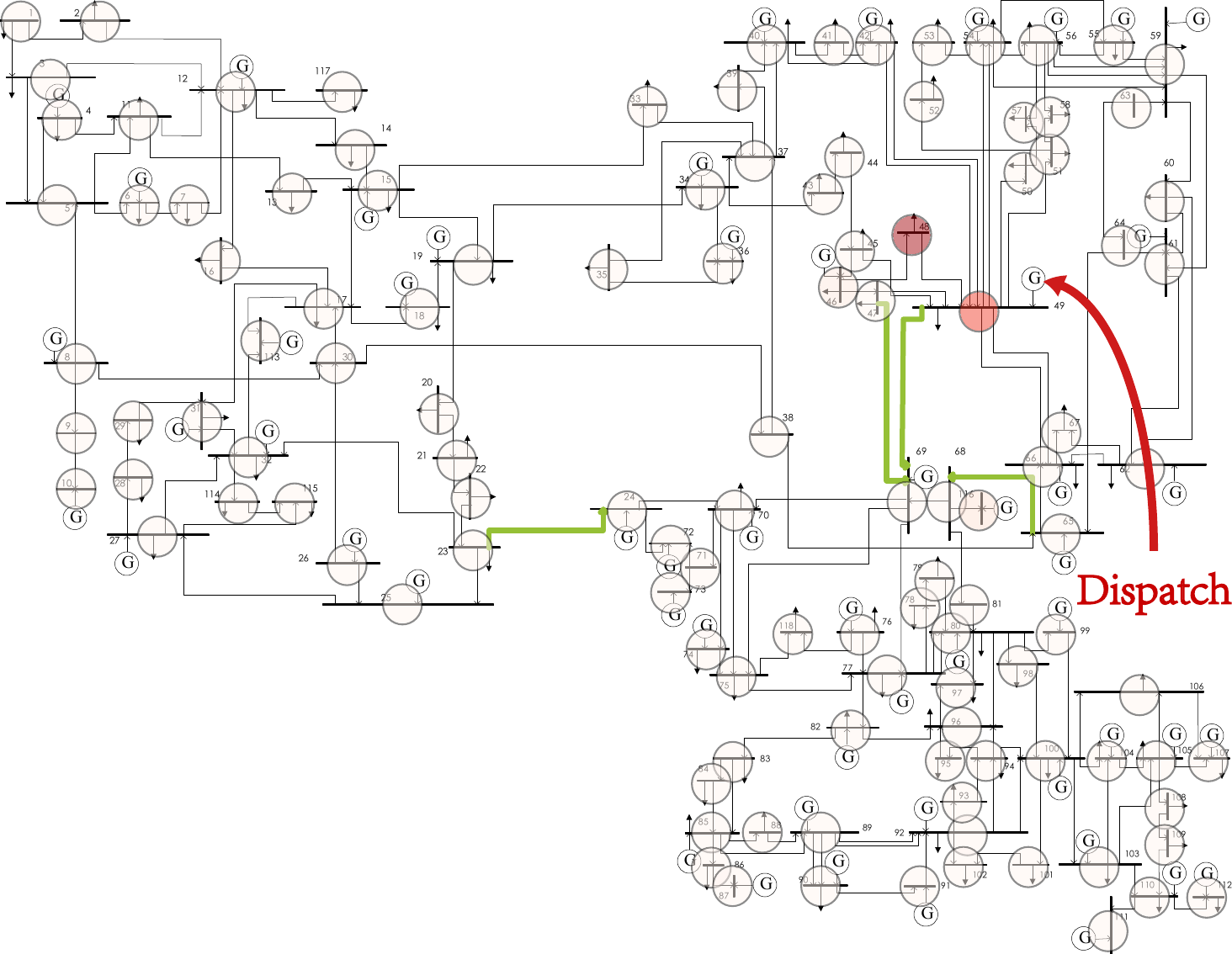}\label{fig:s2_power-0}}\;
  \subfloat{\includegraphics[scale=0.22]{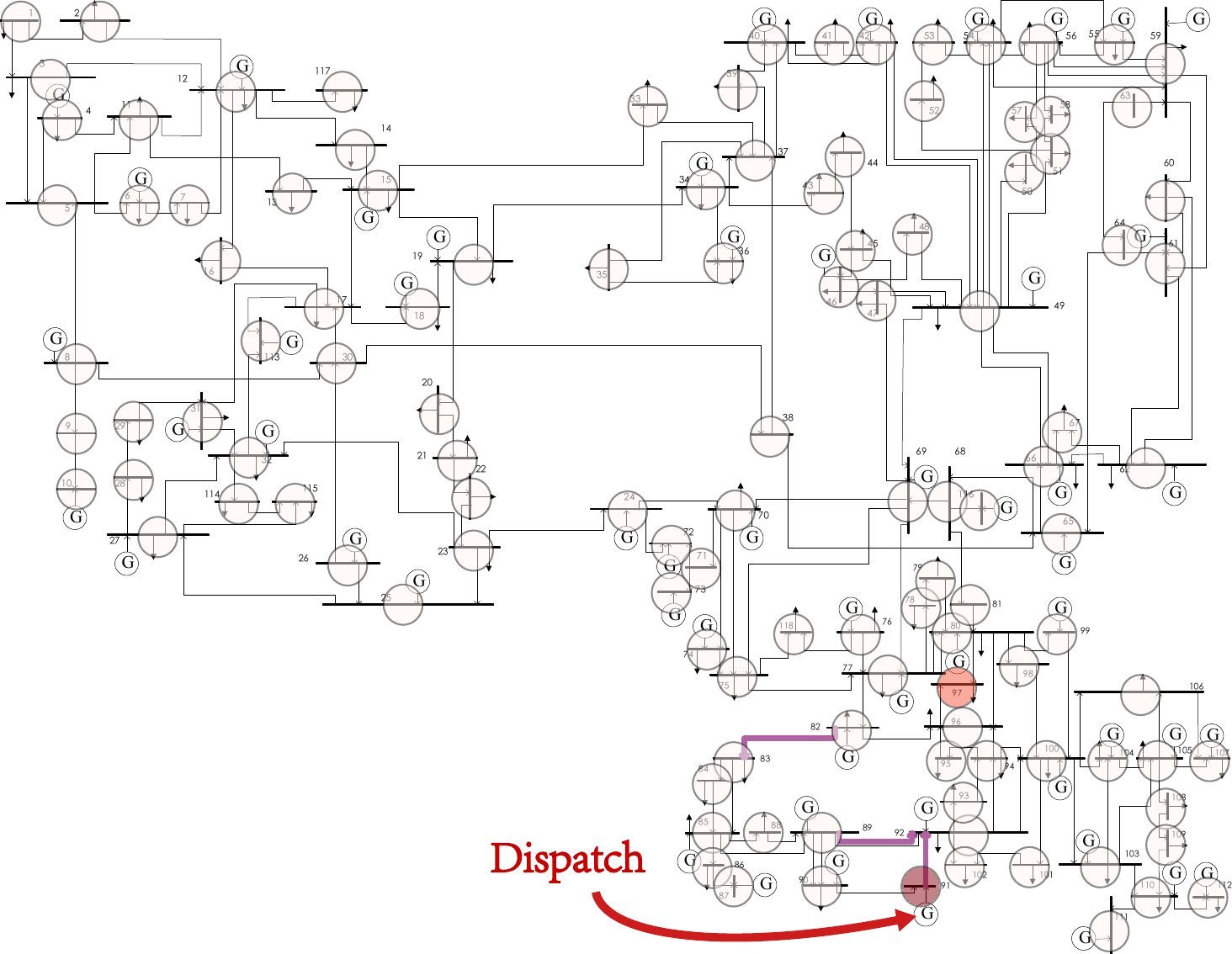}\label{fig:s3_power-0}}\;
  \subfloat{\includegraphics[scale=0.22]{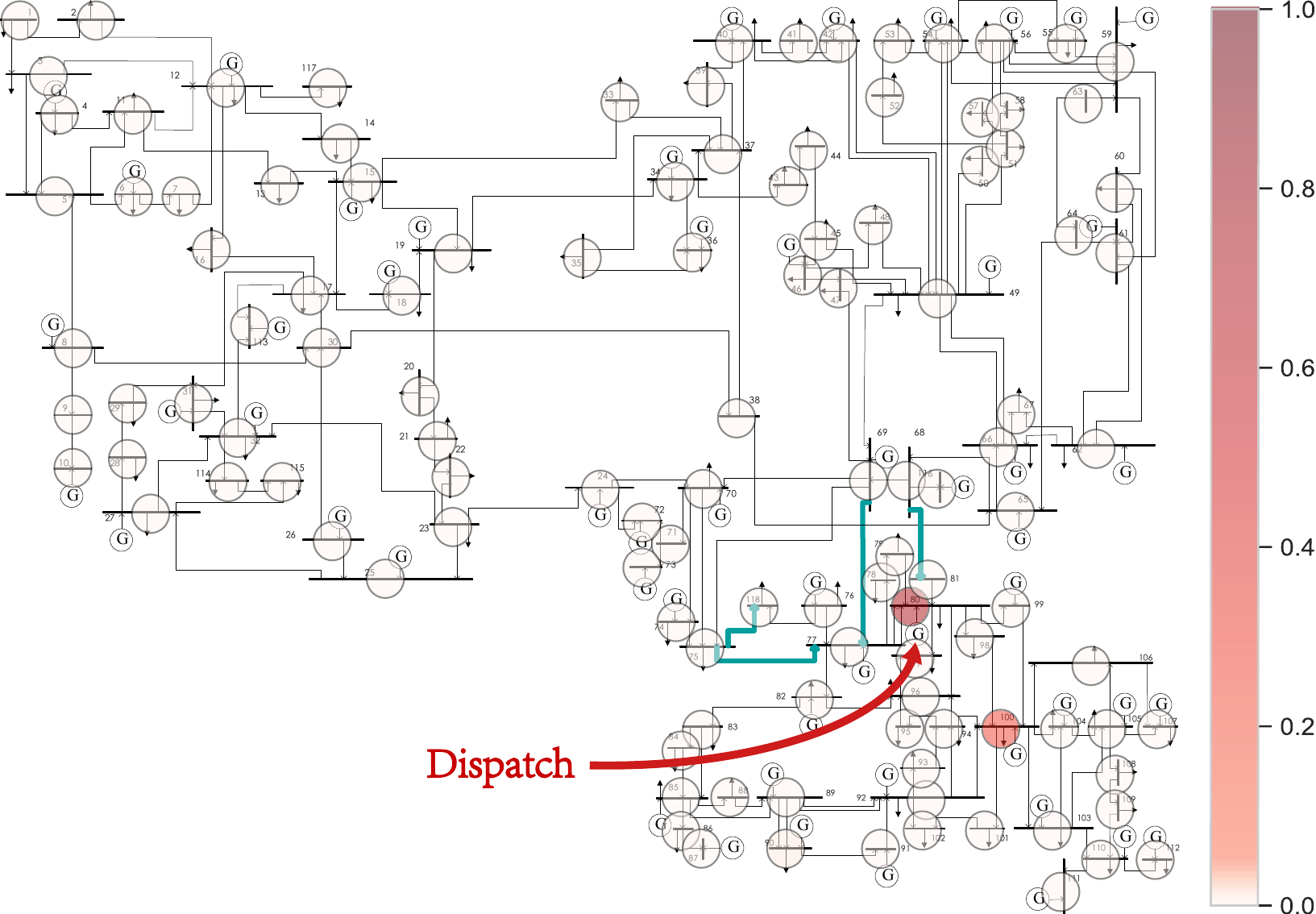}\label{fig:s1_power-0}}
  \caption{A visualization example of the attribution maps in different single-interface scenarios under the IEEE 118-bus system. The attention magnitudes of the node attribution maps are represented by circles with different red shades. Please zoom for better view.}
  \label{fig:vis1}
\end{figure*}

\begin{figure*}[!h]
  \centering
  \subfloat{\includegraphics[scale=0.372]{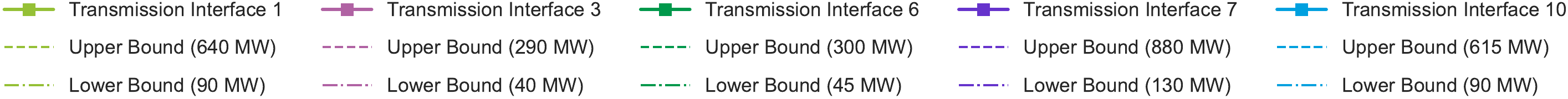}}\\
  \vspace{-0.5em}
  \subfloat{\includegraphics[width=1.0\textwidth]{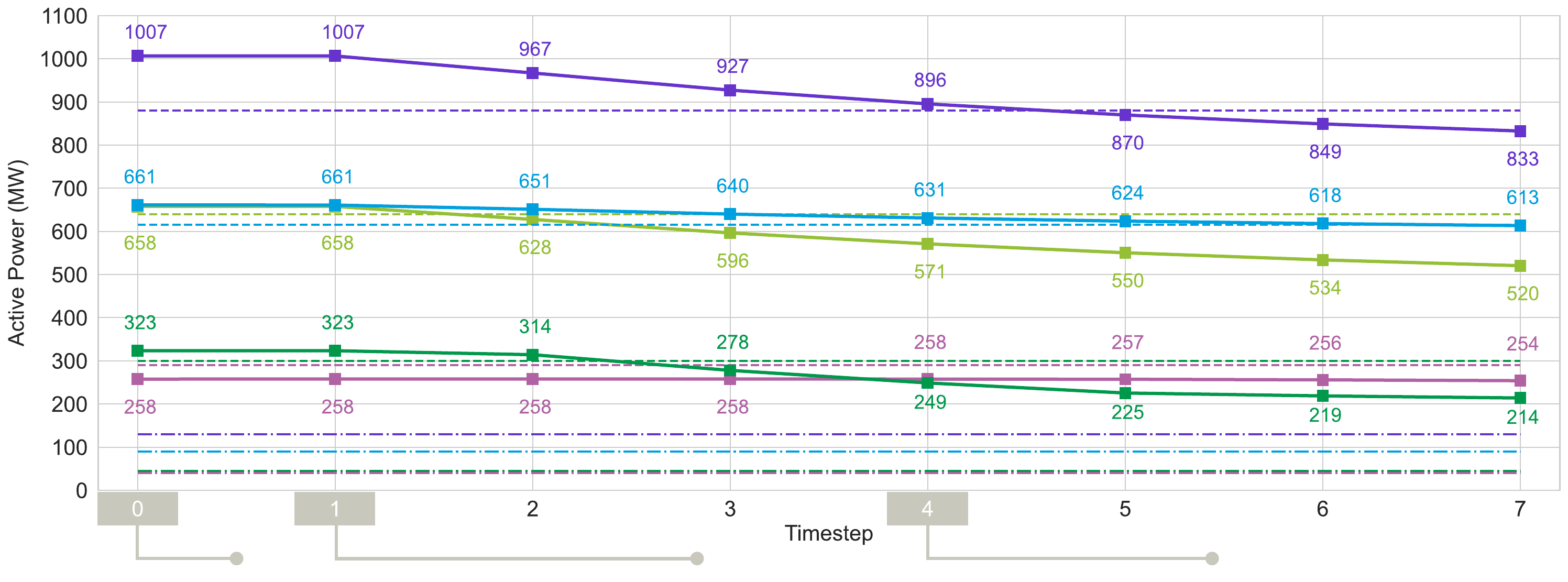}\label{fig:power}}\\
  \vspace{-0.8em}
  \subfloat{\includegraphics[scale=0.22]{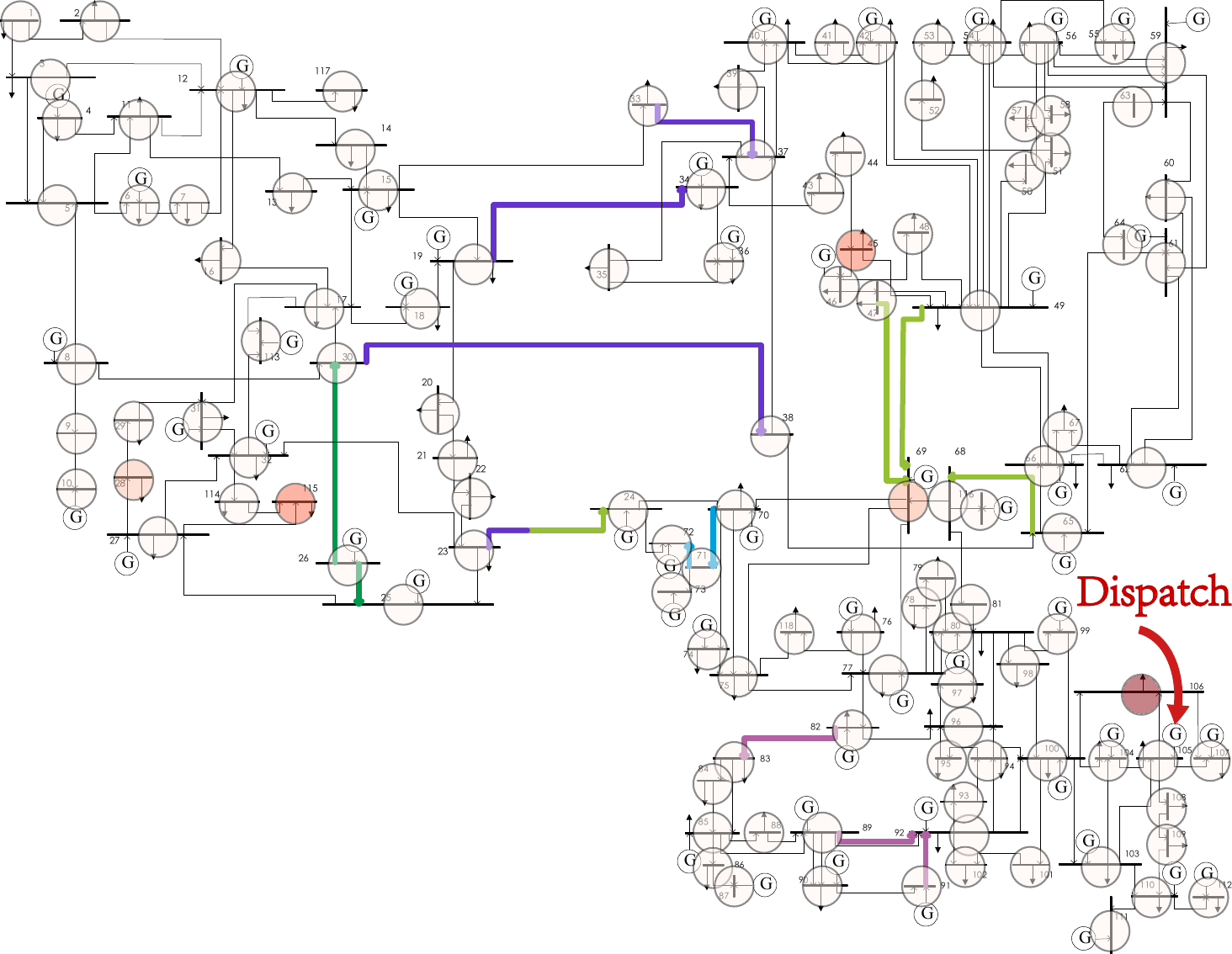}\label{fig:power-0}}\;
  \subfloat{\includegraphics[scale=0.22]{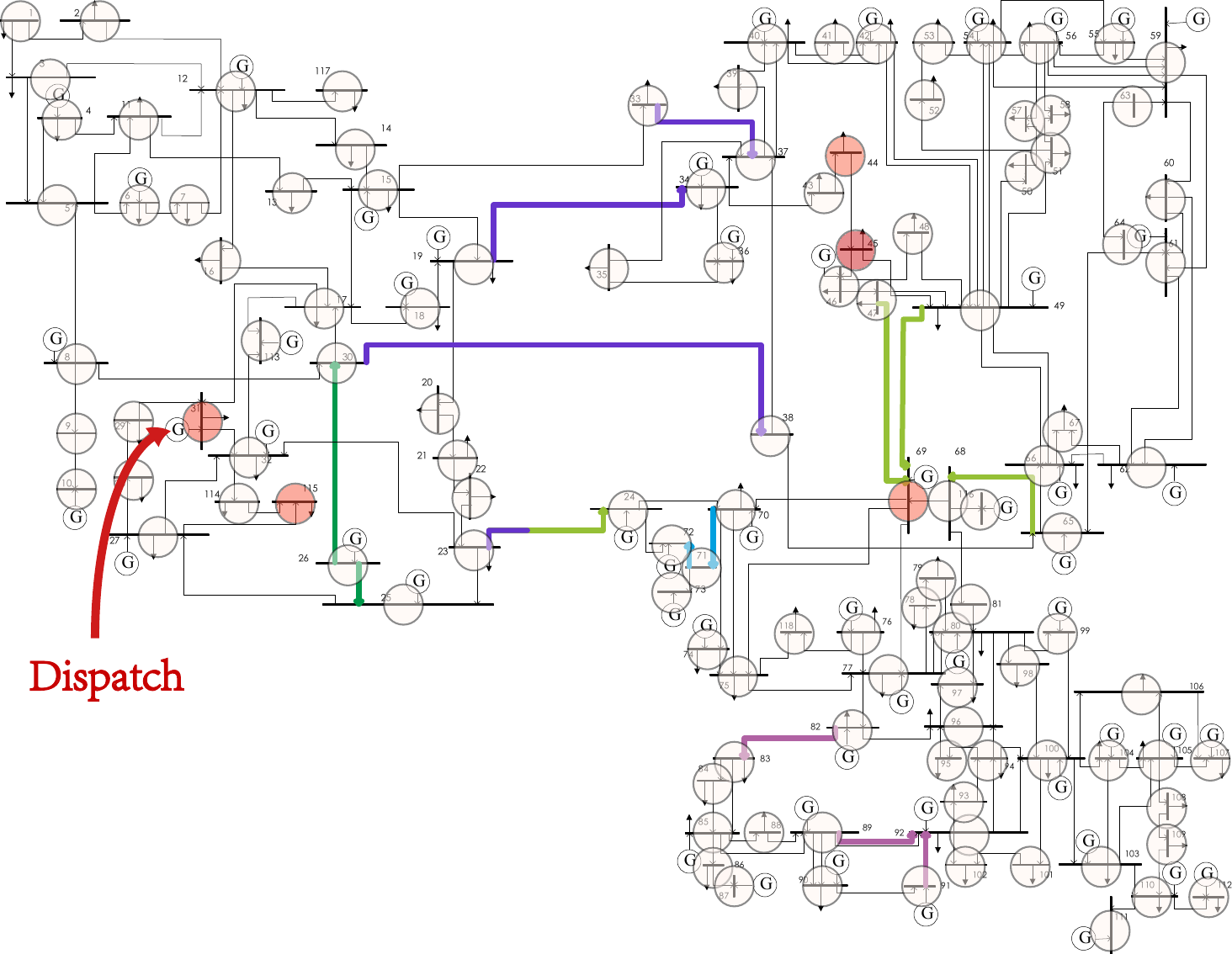}\label{fig:power-1}}\;
  \subfloat{\includegraphics[scale=0.22]{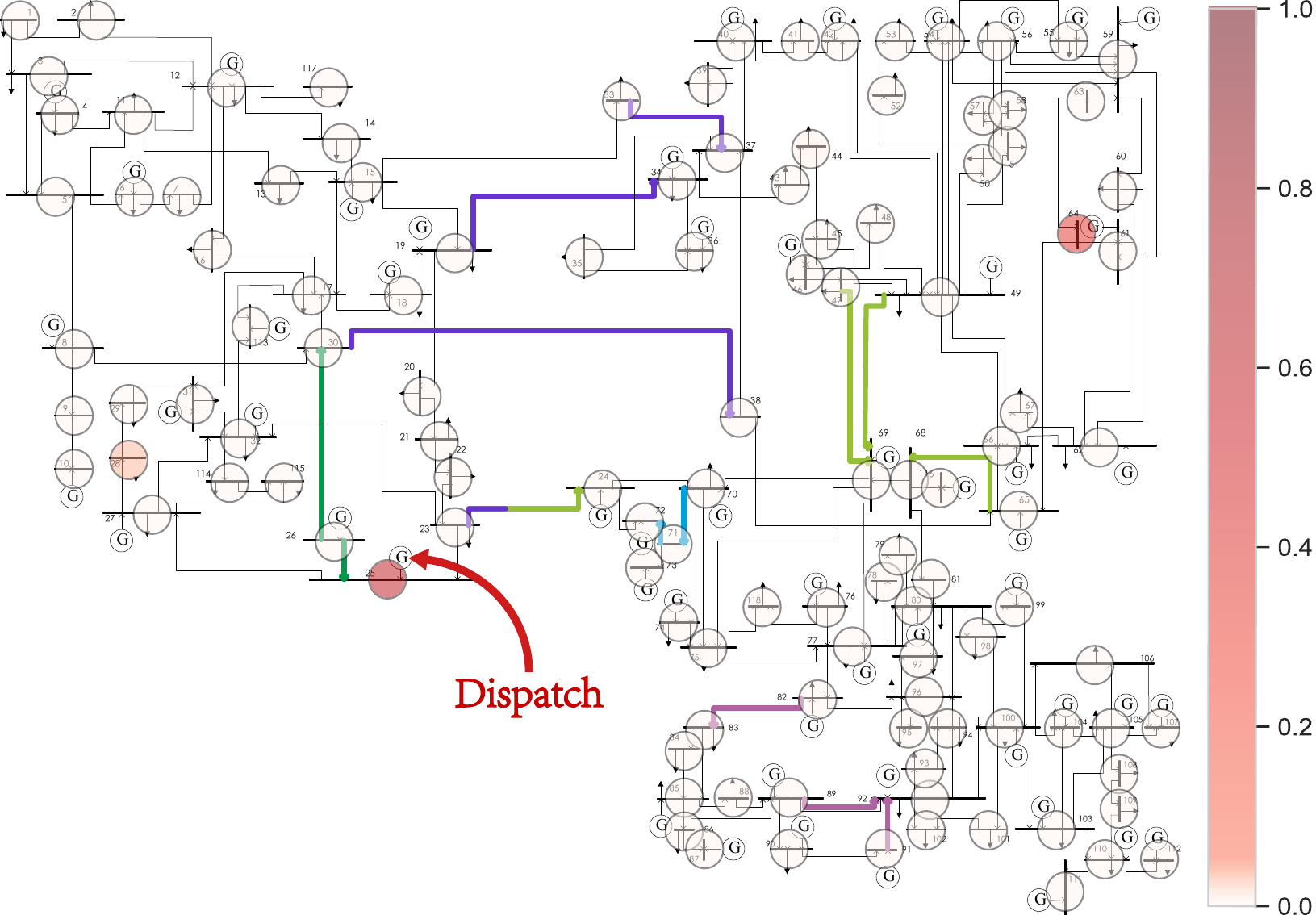}\label{fig:power-2}}
  \caption{A visualization example of the attribution maps in one 5-interface scenario under the IEEE 118-bus system. Please zoom for better view.}
  \label{fig:vis2}
\end{figure*}

\subsection{Ablation Study}
To understand the superior performance of the proposed MAM, we carry out ablation studies to test the contribution of its different components as follows. The comparison results of various ablation on the IEEE 118-bus system with multi-interface tasks are shown in Figure~\ref{fig:ablation} and Table~\ref{tab:ablation}. 
\begin{itemize}[leftmargin=*]
  \item \textbf{MAM-O}. We use one shared GCN instead of two GCN branches to obtain two node-level embedding matrices. The results suggest that sharing parameters can reduce the computation cost while downgrading the final performance.
  This is because two embedding matrices from GCN have different roles in our attribution mechanism.
  One matrix is used to generate the multi-task attribution map, which only distinguishes the impact of different power system nodes on each adjustment task. The other matrix is used to give a compact state representation according to the multi-task attribution map.
  If we only use one shared GCN, it means that two embedding matrices are forced to be the same, which severely damages the model expressiveness.
  \item \textbf{MAM-M}. We directly take the multi-task attribution map instead of the state representation as the input of Dueling Deep Q Networt. By comparing MAM with MAM-M, we can conclude that only multi-task attribution map is not enough to train the subsequent Dueling Deep Q Network. Although the multi-task attribution map enable the agent to pay attention to the important nodes, the agent still fails to make proper decisions and perform poorly due to the lack of state information.
  \item \textbf{MAM-W}. We use the weighted average operation with the learnable parameter instead of the average operation to obtain the task representation for the multi-interface task. The weighted average operation allows the agent to weigh different transmission interfaces rather than regard them as equally important. Thus, it is reasonable that MAM-W can achieve the results on par with and sometimes slightly superior to MAM on the test success rate. However, MAM can obtain the lower test economic cost than MAM-W.
\end{itemize}

\subsection{Visualization Analysis}
To further explain the learned attribution map from MAM, we conduct a qualitative analysis. The attribution maps in different single-interface scenarios under the multi-task setting are visualized in Figure~\ref{fig:vis1}. The simulation results show that the proposed method successfully learns different sparse attribution maps for different transmission interfaces, which significantly reduce the state space. Specially, we find that the high-attention nodes are close to the corresponding transmission interface in electrical distance, which usually has a direct impact on the power flow of the transmission interface. Moreover, the final dispatch generators are also on the nodes with high attention. This crisper example demonstrates the interpretability and effectiveness of the proposed MAM. MAM also demonstrates promising results in the multi-interface scenario as shown in Figure~\ref{fig:vis2}. The learned attribution map in the multi-interface scenario is different from that in the single-interface scenario, which provides the importance nodes for multiple transmission interfaces at once. Then MAM selects the generators to dispatch based on the attribution maps. In general, the visual analysis suggests that the proposed method masters the ability to explicitly distinguish the impact of different power system nodes on each transmission interface. Moreover, MAM also learns the relationship between different transmission interfaces, which enables the generalizable policy to handle the multi-task adjustment problem.

\section{Conclusion\label{sec:conclusion}}
In this work, we propose a novel approach, termed as MAM, that enables us to take advantage of the multi-task attribution map for transmission interface power flow adjustment. MAM follows a restructuring-by-disentangling scheme, where several distinguishable node attentions are generated and then selectively reassembled together for the final focused policy. We validate MAM over the IEEE 118-bus system, a realistic regional system in China, and a very large European 9241-bus system, and showcase that it yields results significantly superior to the state-of-the-art techniques in both single-interface and multi-interface tasks under the multi-task settings. To our best knowledge, this paper is the first attempt towards learning multiple transmission interface power flow~adjustment~tasks~jointly.

\textbf{Limitations.} Currently, we only focus on the generalization over different power flow adjustment tasks under the given transmission interfaces. We can observe that the proposed MAM can successfully generalize to the unseen scenarios and exhibits superior performances. However, MAM cannot directly handle the new transmission interface that the agent has not trained on. The premise of generalization in deep learning is sufficient samples. For example, there may be hundreds of transmission interfaces in IEEE 118-bus system, while we only consider 10 critical transmission interfaces in this work. Thus, it is hard to learn the relationship between the trained transmission interfaces and the new one. It is an interesting direction to study this few-shot generalization~for~our~future~work.

\bibliographystyle{IEEEtran}
\bibliography{ref}

\end{document}